\documentclass[proof]{WileyASNA-v1}
\usepackage{xtab,booktabs}
\articletype{Original Article}%
\received{10 July 2020}
\revised{23 August 2020}
\accepted{23 August 2020}
\doi{10.1002/asna.20210018}
\begin{document}

\title{Tracing the Local Volume galaxy halo-to-stellar mass ratio with satellite kinematics}
\author[1]{Igor Karachentsev}
\author[1]{Olga Kashibadze}
\authormark{Karachentsev \& Kashibadze}
\address{\orgdiv{Special Astrophysical Observatory},\\
         \orgname{Russian Academy of Sciences}, \orgaddress{\country{Russia}}}
\corres{I.\,D.\,Karachentsev, SAO RAS,\\
         Nizhny Arkhyz, Karachay-Cherkessia, 369167 Russia. \email{ikar@sao.ru}}

\abstract{Rapid advance has been made recently in accurate distance measurements
for nearby ($D < 11$~Mpc) galaxies based on the magnitude of the tip of red
giant branch stars resolved with the Hubble Space Telescope. We use
observational properties of galaxies presented in the last version of Updated
Nearby Galaxy Catalog to derive a halo mass of luminous galaxies via orbital
motion of their companions. Our sample contains 298 assumed satellites with
known radial velocities around 25 Milky Way-like massive galaxies and 65 assumed
satellites around 47 fainter dominant galaxies. The average total
mass-to-$K$-band luminosity ratio is $31\pm6 M_\odot/L_\odot$ for the luminous
galaxies, increasing up to $\sim200 M_\odot/L_\odot$ toward dwarfs. The
bulge-dominated luminous galaxies are characterized with
$\langle{}M_T/L_K\rangle = 73\pm15 M_\odot/L_\odot$, while the disc-dominated
spirals have $\langle{}M_T/L_K\rangle = 17.4\pm2.8 M_\odot/L_\odot$. We draw
attention to a particular subsample of luminous spiral galaxies with signs of
declining rotation curve, which have a radial velocity dispersion of satellites
less than 55~km/s and a poor dark matter halo with $\langle{}M_T/L_K\rangle =
5.5\pm1.1 M_\odot/L_\odot$. We note that a fraction of quenched (dSph, dE)
companions around Milky Way-like galaxies decreases with their linear projected
separation as $0.75 \exp(-R_p/350~\mathrm{kpc})$.}



\maketitle


\section{Introduction}\label{sec1}

According to the standard cosmological model ($\Lambda$CDM), most of the mass in
galaxies is concentrated in their dark halos which are by one order of magnitude
more massive than stellar matter. The ratio of dark mass $M_{\rm DM}$ to stellar
mass $M_*$ is different for galaxies with different luminosity. Estimations of
$M_*$ are commonly based on $K_s$ luminosity of galaxies, assuming $M_*/L_K$
ratio to be $1.0M_{\odot}/L_{\odot}$ (Bell et al. 2003) or $0.6 M_{\odot}/
L_{\odot}$ (Lelli et al. 2016). According to Kourkchi \& Tully (2017), the
relation between $\log(M_{\rm DM}/L_K)$ and $\log(L_K)$ has an asymmetrical
V-shaped profile increasing toward reach groups and clusters at $\log(L_K/
L_{\odot})>10.7$ and toward dwarf galaxies at $\log(L_K/L_{\odot}) <9.0$. In the
intermediate luminosity zone the characteristic $M_{\rm DM}/L_K$ ratio has the
minimal value of $\sim 30M_{\odot}/L_{\odot}$. Such a behaviour of $M_{\rm
DM}/M_*$ vs. $M_*$ is qualitatively consistent with the results of N-body
simulations (Sales et al. 2013, Moster et al. 2013, Wechsler \& Tinker, 2018).
The mass of dark matter in a galaxy cluster is usually determined by virial
motions of its members or from the weak lensing effect. Estimations of the full
mass of dwarf galaxies are based on the dispersion of radial velocities of stars
or few globular clusters (if any).

The relation between $M_{\rm DM}$ and $M_*$ (or $L_K$) for masses comparable
with the stellar mass of the Milky Way is still outlined rather unreliably. The
determination of the total mass of a galaxy is most commonly based on data on
radial velocities and projected separations of its satellites. Notably, it works
for a case of a dynamically isolated galaxy, so the motions of its tiny
companions are expected to be Keplerian. The most suitable objects for
determining $M_{\rm DM}$ are nearby bright galaxies surrounded by a retinue of
dwarf satellites. In the Local Volume within 11 Mpc there are about 1000 dwarf
galaxies most of which have measured radial velocities. Their velocities and
projected separations relative to bright galaxies were used to estimate the
total masses of galaxies similar to Milky Way (Karachentsev 2005, Karachentsev
\& Kudrya 2014). Over the last two decades the Hubble Space Telescope performed
wholesale measurements of distances to the Local Volume galaxies from
luminosities of the tip of red giant branch stars (TRGB) with an accuracy of
$\sim5$\%. It allowed to confidently distinguish between physical satellites of
bright galaxies and background or foreground objects. The most full compilation
of individual distances of the Local Volume galaxies is presented in the Updated
Nearby Galaxy Catalog = UNGC (Karachentsev et al.
2013)\footnote{\url{http://www.sao.ru/lv/lvgdb}} and Extragalactic Distance
Database = EDD (Anand et al. 2021)\footnote{\url{http://edd.ifa.hawaii.edu}}.
The UNGC catalog contains data on radial velocities, morphological types,
$K$-luminosities and other parameters of galaxies located at distances
$D<11$~Mpc.

In this work we use the most presentable sample of dwarfs swarming around the
Local Volume galaxies to determine the total (orbital) mass of galaxies of
different luminosity. In Section 2 we present the list of neighboring systems
pooled in stacked groups by the luminosity of the central galaxy in different
ranges. In Section 3 we consider the relation between the mean ratio $M_T/L_K$
and several parameters of galaxies in groups. In Section 4 a special case of
bright galaxies with scanty dark halos is discussed. In Section 5 some
properties of early and late type satellites are considered. In the last section
we give the concluding remarks. Appendix contains the data on 380 companions of
the 23 most bright galaxies of the Local Volume besides the Local Group.

\section{Stacked groups in the Local volume}\label{sec2}

Every galaxy in the UNGC catalog has a dimensionless parameter
$$\Theta_1=\max(\log(L_i/D_i^3)]+C. \,\,\, i= 1, 2, \ldots N,$$
which is termed the {\itshape tidal index}. Here $L_i$ is the luminosity of
neighbouring galaxy in the $K$ band and $D_i$ is the distance between the
considered galaxy and the neighboring galaxy. Ranging neighbours by the tidal
force $F_i\sim L_i/D_i^3$ allows to determine the most significant neighbour,
the so-called {\itshape main disturber} (MD) which name is also noted in the
UNGC for every galaxy. The $C$ constant is adopted in such a way that a galaxy
with $\Theta_1=0$ is situated at the {\itshape zero velocity sphere} relative to
its main disturber. Hence, galaxies with $\Theta_1<0$ correspond to the objects
of the cosmic expansion field while galaxies with $\Theta_1>0$ and the same MD
comprise satellite families around their main disturbers. In certain cases of
hierarchical structure, the main disturber of a galaxy is coincidently a
companion of a more massive galaxy (for example: SMC as a satellite of LMC, and
both Magellanic Clouds are the satellites of the Milky Way). Such a way of
clustering galaxies does not impose constraints on radial velocities of supposed
group members which could lead to selection effects. The imperfection of this
clustering method is the absence of accurate distance estimates for some part of
the Local Volume population.

From the resulting sample of groups of different multiplicity (i.e. families
around the common main disturber), we have selected for our analysis only the
groups with the main galaxy having $K$ luminosity at least twice as high as
luminosity of any of its satellites within the virial radius.

To estimate the total (i.e. {\itshape orbital}) mass of the main galaxy we used
the expression
\begin{equation}
M_{\rm orb}/M_{\odot}=(16/\pi) G^{-1}\langle\Delta V^2
R_p\rangle=1.18\times10^6\langle\Delta V^2 R_p\rangle,
\end{equation}
where $G$ is the gravitational constant, $\Delta V$ is the radial velocity
difference between the satellite and the central galaxy (km~s$^{-1}$), and $R_p$
is the projected separation of the satellite in kpc (Karachentsev \& Kudrya
2014). This relation is based on the assumption that orbits of satellites are
oriented randomly and their typical eccentricity is close to $e=0.7$ according
to the numerical simulations (Barber et al. 2014). However, the observational
data show that in several nearby groups some part of companions is concentrated
into thin planar structures (Ibata et al. 2014, M\"{u}ller et al. 2018). 

Retinues of massive galaxies substantially differ by the number of satellites.
According to Karachentsev et al. (2014) the distribution of the parent galaxies
by the number of their companions $n_{\rm sat}$ can be approximated by equation
\begin{equation}
N(n_{\rm sat})\sim (1+n_{\rm sat})^{-2}.
\end{equation}
Thereby, the abundance of satellites increases from dwarf galaxies to massive
ones. According to our data presented in tables below, the mean number of
companions with luminosity above $10^7L_{\odot}$ depends on $K$ luminosity of
the main galaxy as
\begin{equation}
n_{\rm sat}=(L_K/10^{8.8}L_{\odot})^{1/2}.
\end{equation}

\begin{center}
\begin{table*}[t]%
\caption{Luminous Milky Way-like galaxies in the LV with $\log(L_K/L_{\odot}) > 10.50$.\label{tab1}}
\centering
\begin{tabular*}{500pt}{@{\extracolsep\fill}lrrlrrrccrrrrr@{\extracolsep\fill}}
\toprule
\textbf{Name} & \textbf{T} & \textbf{D} & \textbf{meth} & $\boldsymbol{n_v}$ &
    $\boldsymbol{\langle R_p\rangle}$ & $\boldsymbol{\sigma_v}$ &
    $\boldsymbol{\log L_K}$ & $\boldsymbol{\log M_T}$ & $\boldsymbol{M_T/L_K}$ &
    $\boldsymbol{n_7}$ & $\boldsymbol{n_7e}$ & $\boldsymbol{\Theta_j}$ &
    $\boldsymbol{\log L_1/L_2}$ \\
(1) &(2)& (3) & (4) &(5) & (6) & (7) & (8) & (9) & (10) & (11)&(12) &(13) & (14)\\
\midrule
Milky Way &    4 &  0.01 & geom & 45 & 154 & 109  & 10.70 & 12.07  &  23\,$\pm$\, 8 &  6 &  3 &    1.6 & 1.28\\
M\,31     &    3 &  0.77 & cep  & 51 & 198 & 113  & 10.73 & 12.23  &  31\,$\pm$\, 6 & 15 & 10 &    1.4 & 1.11\\
NGC\,253  &    5 &  3.70 & trgb &  7 & 465 &  42  & 10.98 & 11.91  &   9\,$\pm$\, 3 &  6 &  2 &    0.7 & 1.48\\
NGC\,628  &    5 & 10.19 & trgb &  9 & 279 &  69  & 10.60 & 12.16  &  36\,$\pm$\,13 &  9 &  0 & $-$0.5 & 2.06\\
NGC\,891  &    3 &  9.95 & trgb &  5 & 325 &  92  & 10.98 & 11.92  &   9\,$\pm$\, 2 &  5 &  0 & $-$0.1 & 1.01\\
NGC\,1291 &    1 &  9.08 & trgb &  2 & 254 & 121  & 10.97 & 12.64  &  47\,$\pm$\, 4 & 13 & 10 &    0.2 & 1.67\\
IC\,342   &    6 &  3.28 & cep  &  8 & 282 &  73  & 10.60 & 12.20  &  40\,$\pm$\,20 &  7 &  0 &    0.5 & 1.47\\
NGC\,2683 &    3 &  9.82 & trgb &  2 &  49 &  43  & 10.81 & 11.09  &   2\,$\pm$\, 2 &  3 &  2 & $-$1.2 & 2.95\\
NGC\,2784 & $-$2 &  9.82 & sbf  &  5 & 331 & 146  & 10.80 & 12.96  & 145\,$\pm$\,50 & 20 & 12 &    1.0 & 1.72\\
NGC\,2903 &    4 &  9.17 & trgb &  5 & 295 &  41  & 10.85 & 11.65  &   6\,$\pm$\, 5 &  4 &  1 & $-$0.8 & 2.48\\
M\,81     &    3 &  3.70 & trgb & 31 & 219 & 123  & 10.95 & 12.49  &  35\,$\pm$\,10 & 24 & 10 &    1.5 & 0.36\\
NGC\,3115 & $-$1 &  9.68 & sbf  &  8 & 280 & 112  & 10.95 & 12.62  &  47\,$\pm$\,19 &  9 &  3 &    0.2 & 1.68\\
NGC\,3184 &    6 & 11.12 & SN   &  2 & 268 &  59  & 10.52 & 11.82  &  20\,$\pm$\, 2 &  3 &  1 & $-$0.6 & 2.20\\
NGC\,3521 &    4 & 10.70 & TF   &  3 & 216 &  55  & 11.09 & 11.95  &   7\,$\pm$\, 3 &  6 &  2 & $-$0.2 & 2.47\\
NGC\,3556 &    6 &  9.90 & TF   &  1 & 202 &  89: & 10.52 & 12.27: &      ---     &  1 &  0 & $-$0.3 & 2.85\\
NGC\,4258 &    4 &  7.66 & trgb & 10 & 268 &  96  & 10.92 & 12.32  &  25\,$\pm$\,12 & 16 &  7 &    1.1 & 1.45\\
NGC\,4594 &    1 &  9.55 & trgb & 15 & 431 & 204  & 11.32 & 13.19  &  74\,$\pm$\,24 & 23 & 14 &    0.2 & 2.60\\
NGC\,4736 &    2 &  4.41 & trgb & 15 & 462 &  68  & 10.56 & 12.38  &  66\,$\pm$\,22 & 13 &  1 &    0.8 & 2.37\\
NGC\,5055 &    4 &  9.04 & trgb &  4 & 216 &  54  & 11.00 & 11.71  &   5\,$\pm$\, 2 &  7 &  1 & $-$0.4 & 2.54\\
NGC\,5128 & $-$2 &  3.68 & trgb & 34 & 337 & 124  & 10.89 & 12.67  &  60\,$\pm$\,18 & 33 & 23 &    1.5 & 1.14\\
NGC\,5194 &    5 &  8.40 & SN   &  5 & 269 &  83  & 10.97 & 12.13  &  14\,$\pm$\,12 &  8 &  3 &    1.3 & 0.38\\
NGC\,5236 &    5 &  4.90 & trgb & 10 & 294 &  61  & 10.86 & 12.03  &  15\,$\pm$\, 4 & 18 & 10 &    0.6 & 1.98\\
M\,101    &    6 &  6.95 & trgb &  8 & 281 &  69  & 10.79 & 12.03  &  17\,$\pm$\, 7 &  8 &  2 &    0.1 & 1.59\\
NGC\,6744 &    4 &  9.51 & trgb &  5 & 346 &  71  & 10.91 & 12.19  &  19\,$\pm$\,11 &  9 &  1 &    1.1 & 2.60\\
NGC\,6946 &    6 &  7.73 & trgb &  8 & 323 &  65  & 10.99 & 12.09  &  13\,$\pm$\, 5 &  8 &  0 &    0.6 & 2.59\\
\bottomrule
\end{tabular*}
\end{table*}
\end{center}

For example, for a galaxy like Milky Way with $L_K= 5\times 10^{10} L_{\odot}$
the expected number of satellites (brighter than the stated threshold) is
$n_{\rm sat} = 9$, and for dwarf galaxies with $L_K= 1\times10^8L_{\odot}$ there
is only one expected companion for 2--3 galaxies. The empirical relation (3)
well agrees with the data obtained by Enger et al (2021) in the TNG50 N-body
simulations. Besides that, for early type galaxies (E, S0, Sa) the abundance of
observable satellites is greater than about twice the value for late type
galaxies. The latter was noted by Javanmardi \& Kroupa (2020).

To reduce fluctuation due to the low statistic, we have pooled retinues of
galaxies in stacked groups within several ranges by their $K$ luminosity.
Table~\ref{tab1} presents the basic parameters of 25 Local Volume groups with
main galaxies which luminosities are comparable with the luminosity of the Milky
Way and exceeds the threshold of $L_K/L_{\odot}=10.50$ dex. Its columns contain:
(1) galaxy name; (2) de Vaucouleurs morphological type; (3) galaxy distance in
Mpc; (4) method applied to estimate distance: {\itshape cep}--from cepheids
luminosity, {\itshape trgb}--from the tip of the red giant branch, {\itshape
sbf}--from surface brightness fluctuation, {\itshape SN}--from supernovae
luminosity, {\itshape TF}--from Tully \& Fisher relation between the 21~cm line
width and the luminosity of a galaxy; (5) number of satellites with radial
velocity estimations; (6) mean projected distance of satellites from the main
galaxy in kpc; (7) rms radial velocity of satellites relative to the main galaxy
(in km~s$^{-1}$); (8) luminosity of the main galaxy in $K$ band from the UNGC
catalog in solar luminosity units; (9) total (orbital) mass estimate for the
main galaxy in solar mass units; (10) total mass-to-$K$ luminosity ratio of the
galaxy and its statistical error; (11, 12) number of expected satellites with
luminosity $L_K/L_{\odot}>7.0$ dex and number of early type satellites (dSph, E)
among them; (13) logarithmic average $K$ luminosity density within 1 Mpc around
the considered galaxy, $\Theta_j$, in global mean density units
$4.28\times10^8L_{\odot}/{\rm Mpc}^3$ (Jones et al. 2006); (14) logarithmic
ratio between luminosity of the main galaxy and luminosity of its brightest
companion situated within the virial zone. As it can be seen from the last
column data, nearly all the main galaxies, except M81 and NGC5194, are an order
of magnitude or even more brighter than their satellites, corresponding to the
model of low-mass test particles moving around the massive central body.

The data on distances, radial velocities and luminosities of 380 supposed
companions around the 23 most bright galaxies of the Local Volume besides the
Local Group are presented in Appendix Table~\ref{tabapp} (available also
online\footnote{\url{http://cdsarc.u-strasbg.fr/viz-bin/cat/J/AN/...}}). In
addition, we use the properties of 45 Milky Way satellites and 51 M31 satellites
from the sample of Kashibadze \& Karachentsev (2018) and Putman et al. (2021).
The source of extragalactic data is the latest version of the UNGC catalog
(http://www.sao.ru/lv/lvgdb) supplemented by recent publications (Byun et al.
2020, Karachentsev et al. 2020, Karachentsev \& Makarova 2019, M\"{u}ller \&
Jerjen 2020, Anand et al. 2021). The UNGC site contains references to original
publications introducing the data have been used in this paper.

Galaxies less bright than the Milky Way have only a few satellites. The basic
parameters of these galaxies and their companions are listed in Tables
\ref{tab2}--\ref{tab5}. The objects are stacked by $K$ luminosity ranges of main
galaxies: $[10.0-10.5]$~dex, $[9.5-10.0]$~dex, $[9.0-9.5]$~dex and $<9.0$~dex.
The columns of these tables contain: (1) galaxy name fixed in the UNGC catalog;
(2) numeric morphological type; (3) distance in Mpc; (4) method applied to
estimate distance; satellites without individual distance estimates but supposed
to be group members are noted as {\itshape mem}; one galaxy has distance
estimated from the Numerical Action Method (NAM) model (Shaya et al, 2017); (5)
radial velocity of a galaxy relative to the Local Group centroid in km~s$^{-1}$;
(6) logarithmic $K$ luminosity in solar luminosity units; (7) projected
separation in kpc under the assumption that a satellite has the same distance as
the main galaxy; (8) radial velocity of a satellite relative to the main galaxy.
The tables include also some supposed companions without having yet measured
radial velocities.

As a result, the total number of satellites in our analysis amounts to $N =
564$. Among them 363 companions, or 64\%, have measured radial velocities. As it
can be seen from presented data, many satellites of neighboring galaxies still
miss accurate distance estimates, despite the focused Hubble Space Telescope's
efforts to measure TRGB distances. In that respect the situation beside the
Local Volume seems to be even less provided with observational data.

\section{The total mass-to-luminosity ratio at various parameters of neighboring
groups.}\label{sec3}

Mean characteristics of the main galaxies in the neighboring groups are listed
in Table~\ref{tab6}. Its first column indicates $K$ luminosity range of the
dominating galaxy. In the second column we give number of satellites with
measured radial velocities used to estimate the orbital (total) mass, $M_T$. The
third column contains mean radial velocities of satellites relative to the main
galaxy. The fourth column specifies mean luminosity of the main galaxy in solar
luminosity units. The fifth and the sixth columns are mean projected separation
of satellites (kpc) and  rms radial velocity of satellites (km~s$^{-1}$)
relative to the central galaxy. The last two columns contain mean values of
orbital mass (in $M_{\odot}$) and orbital mass-to-$K$ luminosity ratio for the
main galaxy with mean error estimates.

As it follows from the data listed in the third column, mean radial velocities
of satellites agree with the radial velocity of the galaxy, giving the evidence
of their physical connection. The agreement of radial velocities within
statistical errors is found in every group. The only exception is the NGC2784
group with all 5 satellites having radial velocities exceeding the radial
velocity of the main galaxy. These galaxies does not have accurate distance
estimates, and the $M_T/L_K=145 M_{\odot}/L_{\odot}$ value for this group is
extremely large among the groups of Table~\ref{tab1}.

\begin{figure*}[t]
\centerline{\includegraphics[width=98mm]{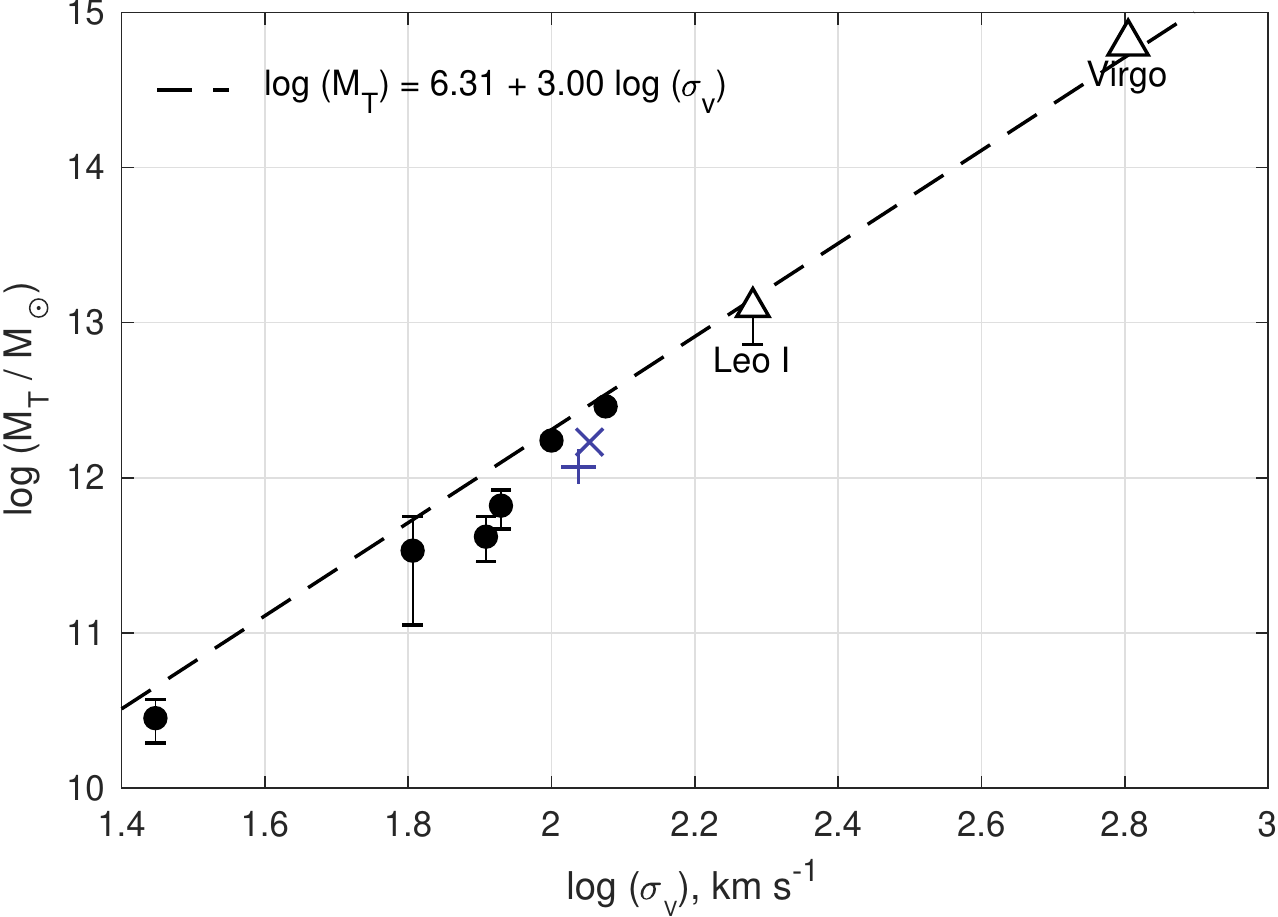}}
\caption{Correlation between virial (orbital) mass and radial velocity
dispersion. The stacked groups in the Local Volume are presented as solid
circles, the Milky Way and M31 galaxies are depicted by plus and cross signs,
respectively. The dashed line corresponds to Tully (2015) relation for rich
groups and clusters.\label{fig1}}
\end{figure*}

According to Tully (2015), the radial velocity dispersion in groups and
clusters, $\sigma_v$, and the virial mass of these systems, $M_T$, follow the
tight relation
\begin{equation}
\log(M_T/M_{\odot})=6.31+3.0\log(\sigma_v) 
\end{equation}
in the mass range $M_T= (2\times10^{12} - 2\times10^{15})M_{\odot}$.
Figure~\ref{fig1} reproduces the relation between $M_T$ and $\sigma_v$ according
to our data. The stacked Local Volume groups from Table~\ref{tab6} are depicted
by solid circles with error bars. Galaxy groups around the Milky Way and M31 are
shown with plus and cross markers, respectively. We have labeled with triangles
the nearest Virgo cluster (Kashibadze et al. 2020) and the rich group LeoI at
the far periphery of the Local Volume. The Leo I group contains several
brightest galaxies of relatively similar luminosity and thus is not included in
Table~\ref{tab1}. The parameters of the Leo I group are extracted from
Karachentsev \& Kudrya (2014) and augmented with new data on radial velocities
and distances of its members. The relation (4) suggested by Tully (2015) is
depicted by dashed line. It agrees quite well with our data and can be
remarkably extended to dwarf galaxies area with masses as low as
$M_T\sim3\times10^{10}M_{\odot}$. The radial velocity dispersion of satellites
around the Milky Way lays rightward of the relation (4), as we observe
predominantly radial motions of satellites from the interior and thereby see
their almost full velocity vectors.

\begin{figure*}[t]
\centerline{\includegraphics[width=101mm]{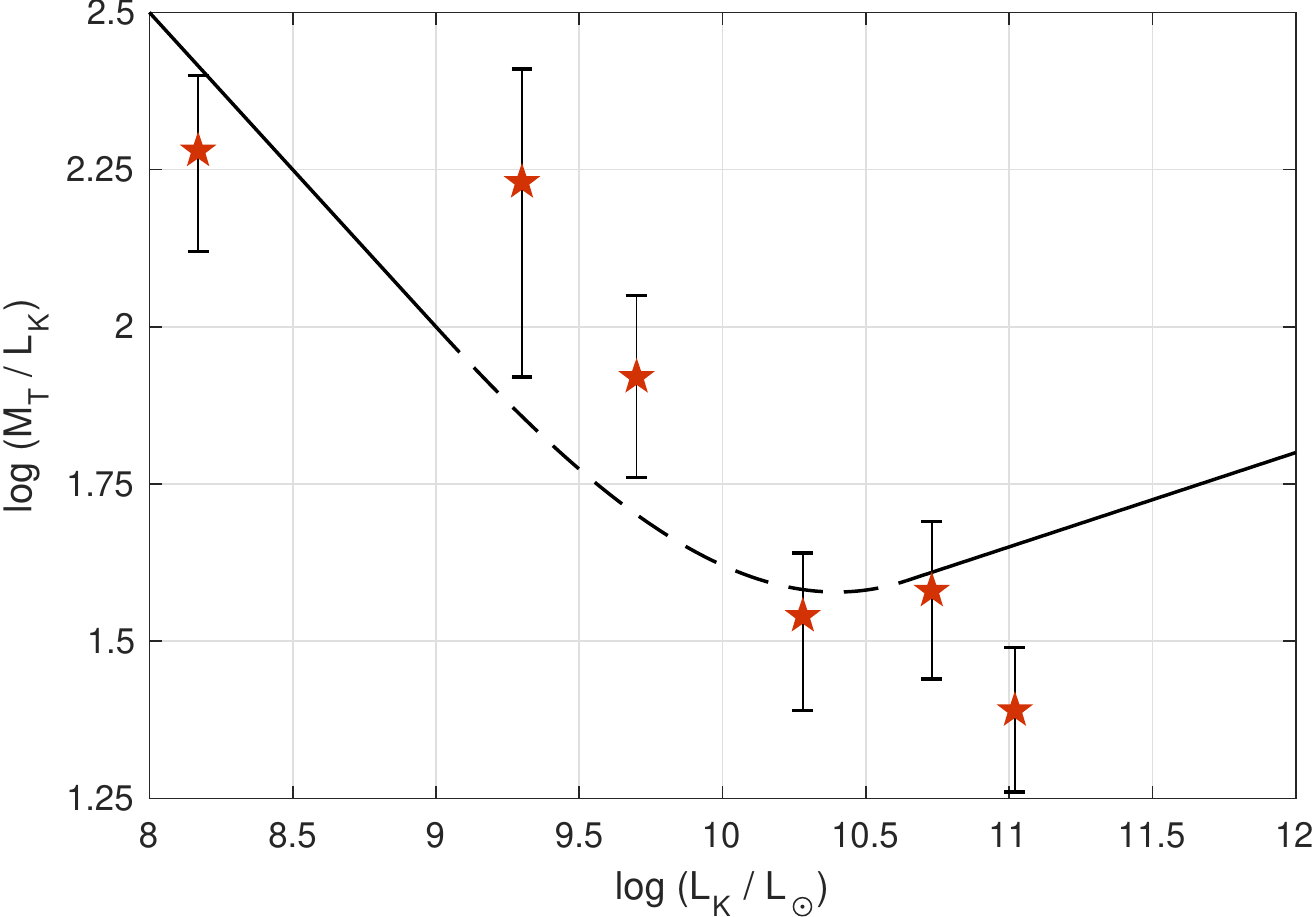}}
\caption{Relation between orbital mass and $K$ luminosity for stacked groups in
the Local Volume. The curved line reflects the relation obtained by Kourkchi
\& Tully (2017).\label{fig2}}
\end{figure*}

Figure~\ref{fig2} reproduces the total mass-to-$K$ luminosity ratio for stacked
groups of the Local Volume as a function of luminosities of groups from the
Table~\ref{tab6}. The curved line on the plot represents Kourkchi \& Tully
(2017) relation:
\begin{equation}
\log(M_T/L_K)=1.50-0.50\times \log(L_K/10^{10}L_{\odot}) \; \; \mbox {at} \; \;\log L_K<9.0
\end{equation}
and
$$\log(M_T/L_K)=1.50+0.15\times \log(L_K/10^{10}L_{\odot}) \; \; \mbox {at} \; \;\log L_K>10.7 $$
with minimum near $L_K/L_{\odot}\simeq10.5$ dex. The stacked groups of the Local
Volume generally follow the relation (5), although the minimal value
$(M_T/L_K)_{\rm min}\simeq25M_{\odot}/L_{\odot}$ turns out to be shifted to
brighter luminosities, $L_K/L_{\odot}\simeq11$ dex. The existence of the minimum
suggests that most star formation activity is inherent in galaxies of this
specific stellar mass. Note that van Uitern et al. (2016) have estimated the
$M_T/M_*$ ratio for Galaxy And Mass Assembly (GAMA) survey objects augmented
with $M_T$ data from Kilo Degree Survey using gravitational lensing effect. They
obtained the minimal value $(M_T/M_*)_{\rm min} =41\pm10$ with Hubble parameter
$H_0=73$~km~s$^{-1}$Mpc$^{-1}$ which is achieved for galaxies with stellar mass
$M_*/M_{\odot}=10.3$ dex. However, Lapi et al. (2018) and Posti \& Fall (2021)
relate the minimum value of $M_T/M_* = 20$ to a stellar mass of $M_*/M_{\odot}=
10.7 dex$. 

The literature consistently notes that the amount of dark matter per unit
stellar mass is much larger in early type galaxies than in those of late types.
Considering the data on radial velocities and projected separations of 214 faint
satellites around 2MASS Isolated Galaxies, Karachentseva et al. (2011) obtained
the median value of $M_T/L_K=64 M_{\odot}/L_{\odot}$ for E and S0 galaxies,
while for spiral galaxies this ratio amounts to $17 M_{\odot}/L_{\odot}$. The
almost four times difference in $M_T/L_K$ is obviously caused by heterogeneous
disk and bulge formation histories. According to Correa \& Schaye (2020), the
diference in $M_T/M_*$ between early and late type galaxies is not so large,
amounting only to factor of 1.4. Whereas More et al. (2011), Mandelbaum et al.
(2016), Posti \& Fall (2021) and Bilicki et al. (2021)  get the value (2-4) for
this ratio. Moreover, Bilicki et al. (2021) note that for high stellar mass (
$M_* > 5 10^{11} M_{\odot}$), the red galaxies occupy dark matter halos that are
much more massive than those occupied by blue galaxies with the same stellar
mass.

\begin{figure}[t]
\centerline{\includegraphics[width=78mm]{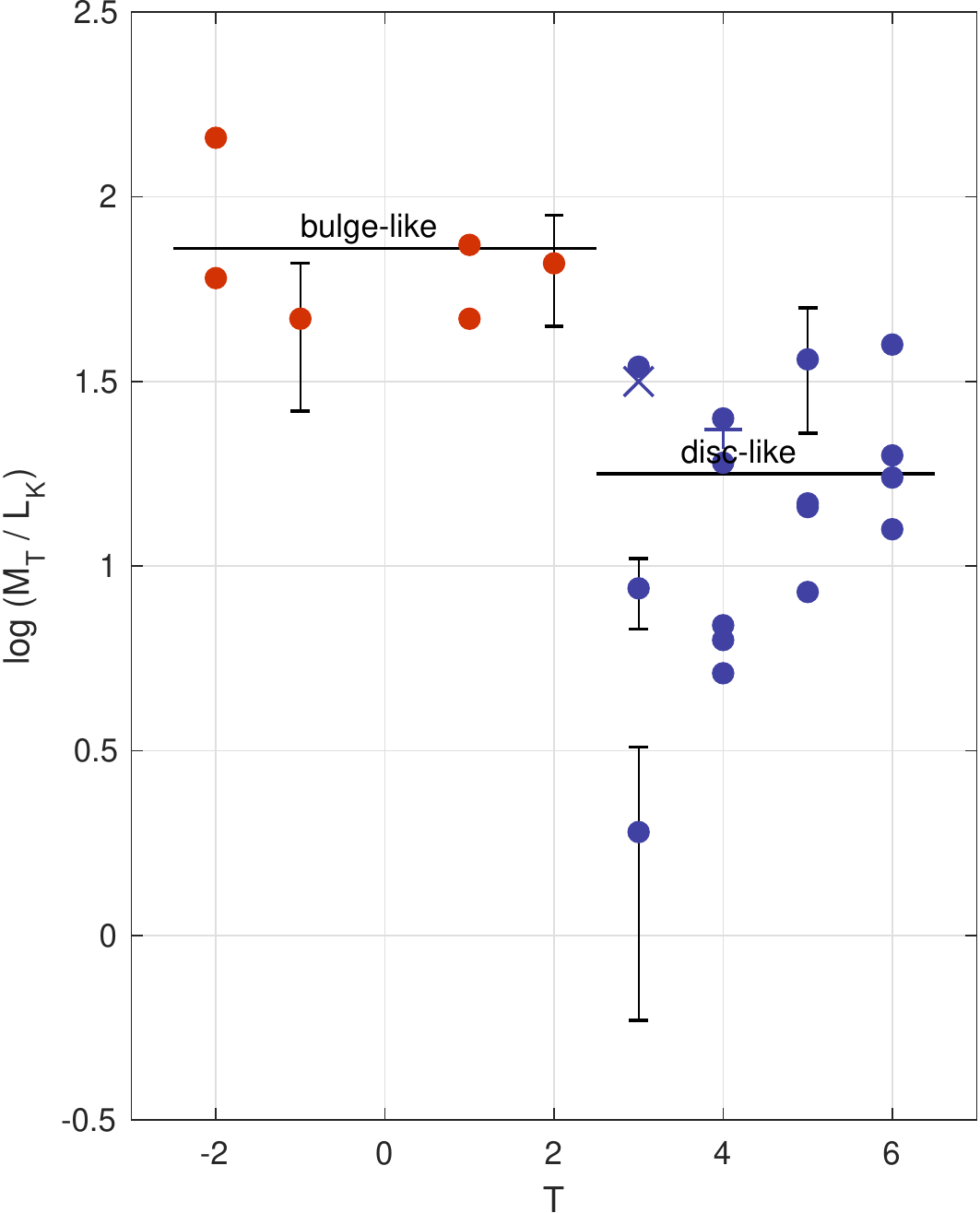}}
\caption{Total mass-to-$K$ luminosity ratio for the Local Volume Milky Way-like
galaxies of different morphological types. The horizontal lines correspond to 
the mean values for the bulge-dominatad and disc-dominated galaxies. The Milky
Way and M31 are depicted by plus and cross signs.\label{fig3}}
\end{figure}

The distribution of bright Local Volume galaxies of different types by their
$M_T/L_K$ ratio is presented in Figure~\ref{fig3}. The vertical bars show
standard $M_T/L_K$ errors for several galaxies from Table~\ref{tab1}. Early type
galaxies with $T<3$, i.e. E, S0, Sa, have the mean ratio $\langle M_T/L_K\rangle
= 73\pm15(M_{\odot}/ L_{\odot})$, while for galaxies with $T\geq3$ dominated by
disks this ratio makes $17.4\pm2.8(M_{\odot}/L_{\odot})$. The difference in
$M_T/L_K$ estimates for bulges and disks attains the factor of $\sim4$,
significantly exceeding the errors of mean. Our Galaxy (plus sign) and the
neighboring galaxy M31 (cross sign) do not seem unique among the disc-like
galaxies. Correa \& Schaye (2020) suggest that disc-like galaxies, as opposed to
bulge-like galaxies with the same halo mass, had more time for gas accretion and
formation of more massive stellar subsystem.

\begin{figure}[t]
\centerline{\includegraphics[width=84mm]{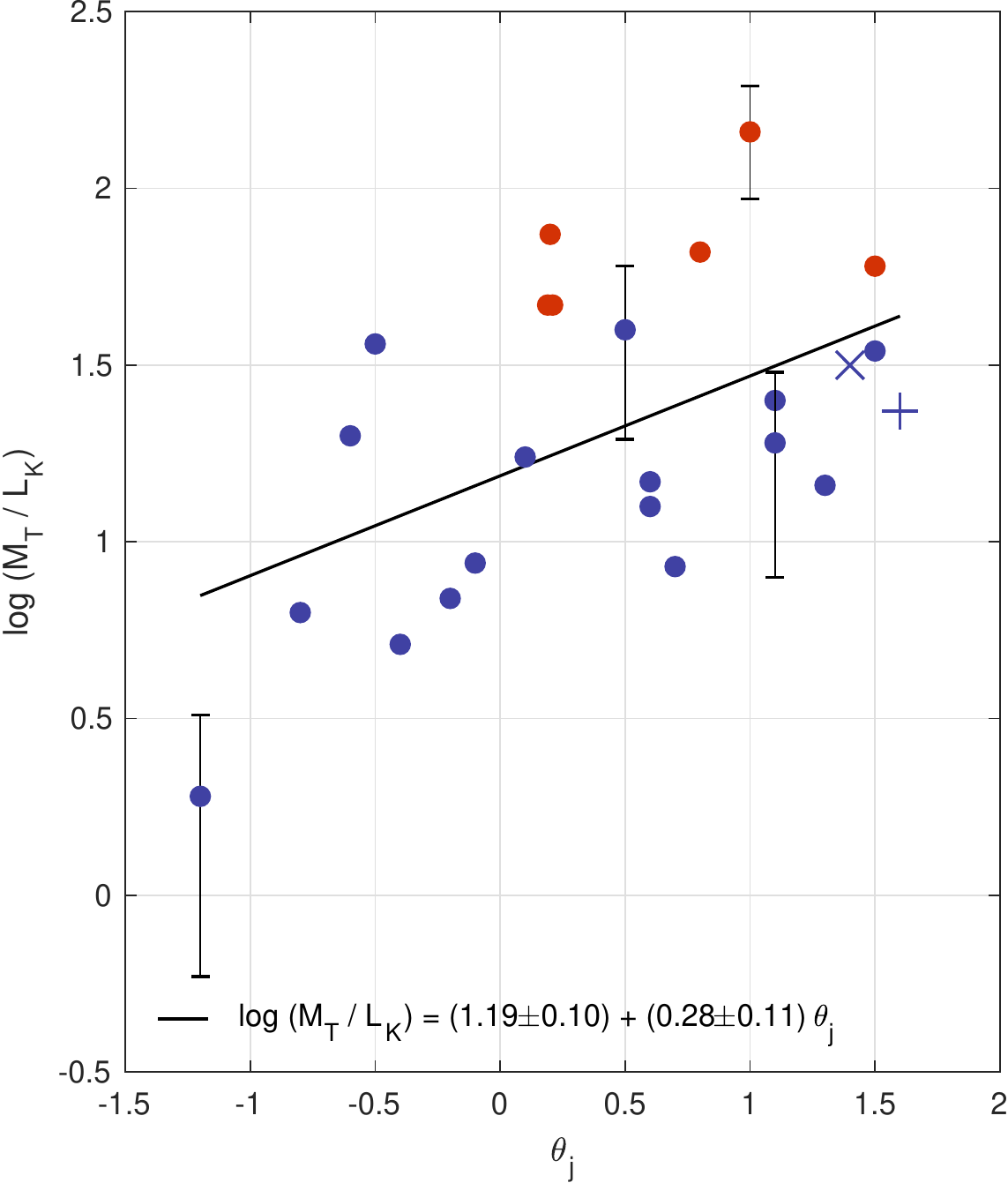}}
\caption{Correlation between total mass-to-$K$ luminosity ratio and logarithmic 
stellar mass density contrast within 1 Mpc around bright galaxies in the Local
Volume. Early type galaxies (E, S0, Sa) are marked by red circles. The Milky Way
and M31 are shown by plus and cross signs.\label{fig4}}
\end{figure}

The idea that the relation between dark and luminous matter can depend not only
on the internal properties of galaxies but also on their environment can be
confirmed by data presented in Figure~\ref{fig4}. The $M_T/L_K$ relation for
galaxies listed in Table~\ref{tab1} is shown as a function of stellar density
contrast inside a 1~Mpc sphere centered on the considered galaxy (in the mean
cosmic density units). Galaxies of late and early types are depicted as blue
and red circles, respectively. It can be seen that early type galaxies are
concentrated in the higher density region, $\Theta_j>0$, having larger $M_T/L_K$
values. In the under-density region, $\Theta_j<0$, there are only disc-dominated
galaxies. Based on the regression line slope, the drop of $M_T/L_K$ ratio from
the richest regions to the sparse ones can achieve the factor of $\sim4$. Our
galaxy (plus sign) and M31 (cross sign) show an unsignificant deviation from the
general trend.

\begin{figure}[t]
\centerline{\includegraphics[width=85mm]{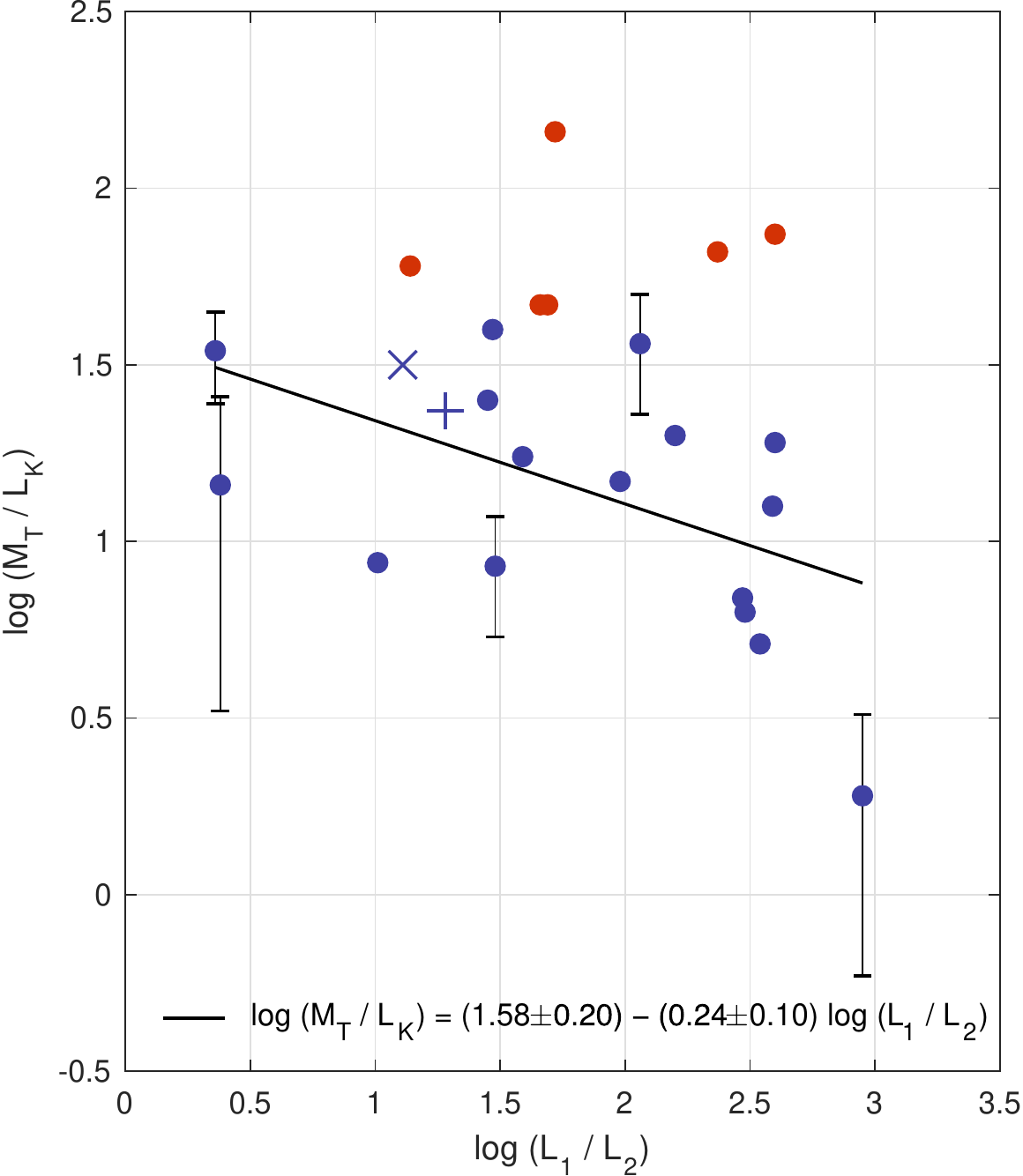}}
\caption{Correlation between the total mass-to-$K$ luminosity ratio and the
ratio of $K$ luminosities of the main galaxy and its brightest satellite. The
designations of galaxies are the same as in Figure~\ref{fig4}.\label{fig5}}
\end{figure}

\begin{figure}[t]
\centerline{\includegraphics[width=85mm]{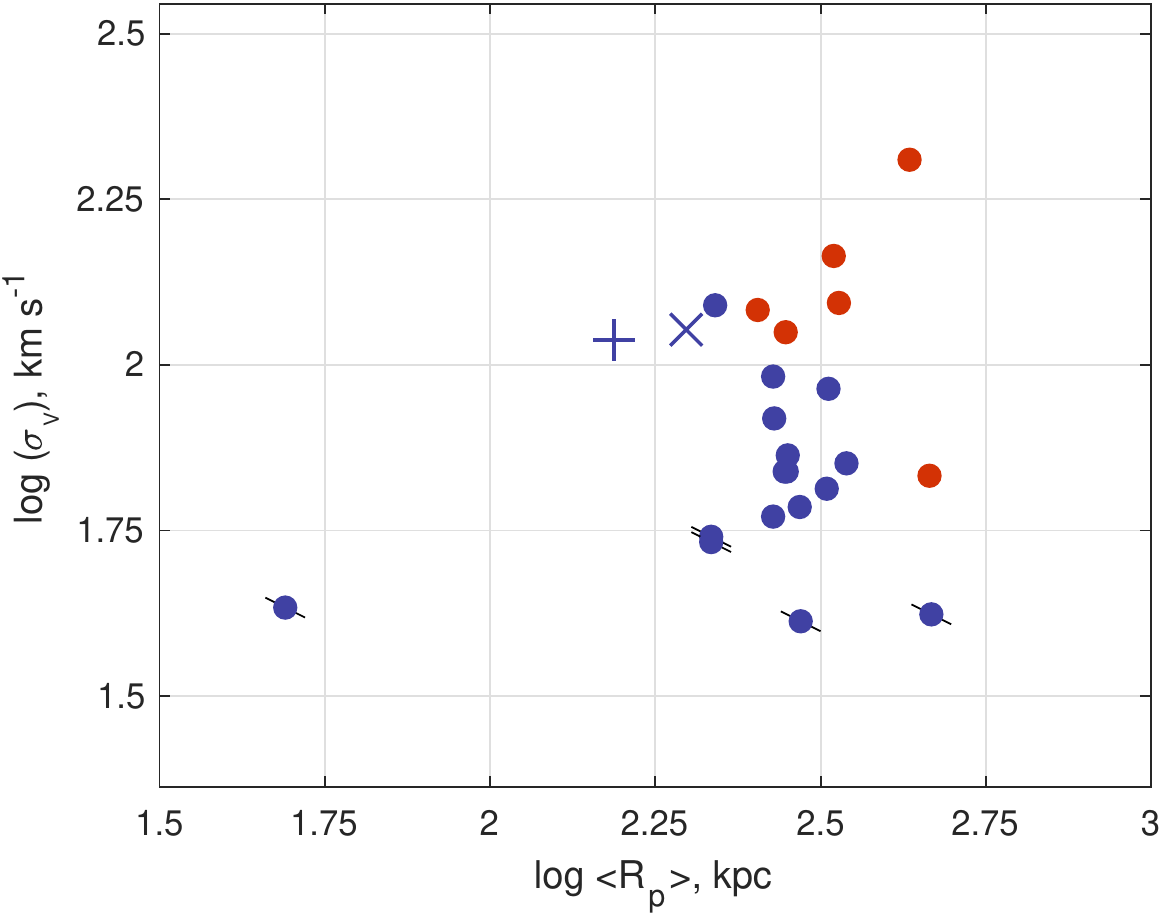}}
\caption{Distribution of the Local Volume bright galaxies by the radial velocity
dispersion and the mean projected separation of satellites. The designations of
galaxies are the same as in previous figure. Galaxies with declining rotation
curves are marked by inclined bars.\label{fig6}}
\end{figure}

Wang et al. (2021) examined the abundance of satellites around isolated central
galaxies in the Sloan Digital Sky Survey and other surveys. The authors
concluded that systems with large difference in luminosities of the central
galaxy and its companions have less massive dark halo. Our data on bright
isolated Local Volume galaxies are consistent with this predication.
Figure~\ref{fig5} reproduces the relationship between $M_T/L_K$ and the ratio of
$K$ luminosities of the main galaxy and its brightest companion located within
the virial zone of the main galaxy. The designations are the same as in
Figure~\ref{fig4}. The regression line $\log(M_T/L_K) = 1.58-0.24\log(L_1/L_2)$
has a negative slope confirming that the dark halo mass density arround the
central galaxy tends to decrease while companions become progressively smaller.
This effect is mostly expressed for late type spirals (blue circles). The Milky
Way and the M31 (plus and cross markers) don't stand out from other bright
galaxies of the Local Volume.

\section{The Milky Way type galaxies with poor dark halos.}\label{sec4}

\begin{figure*}[t]
\centerline{\includegraphics[width=116mm]{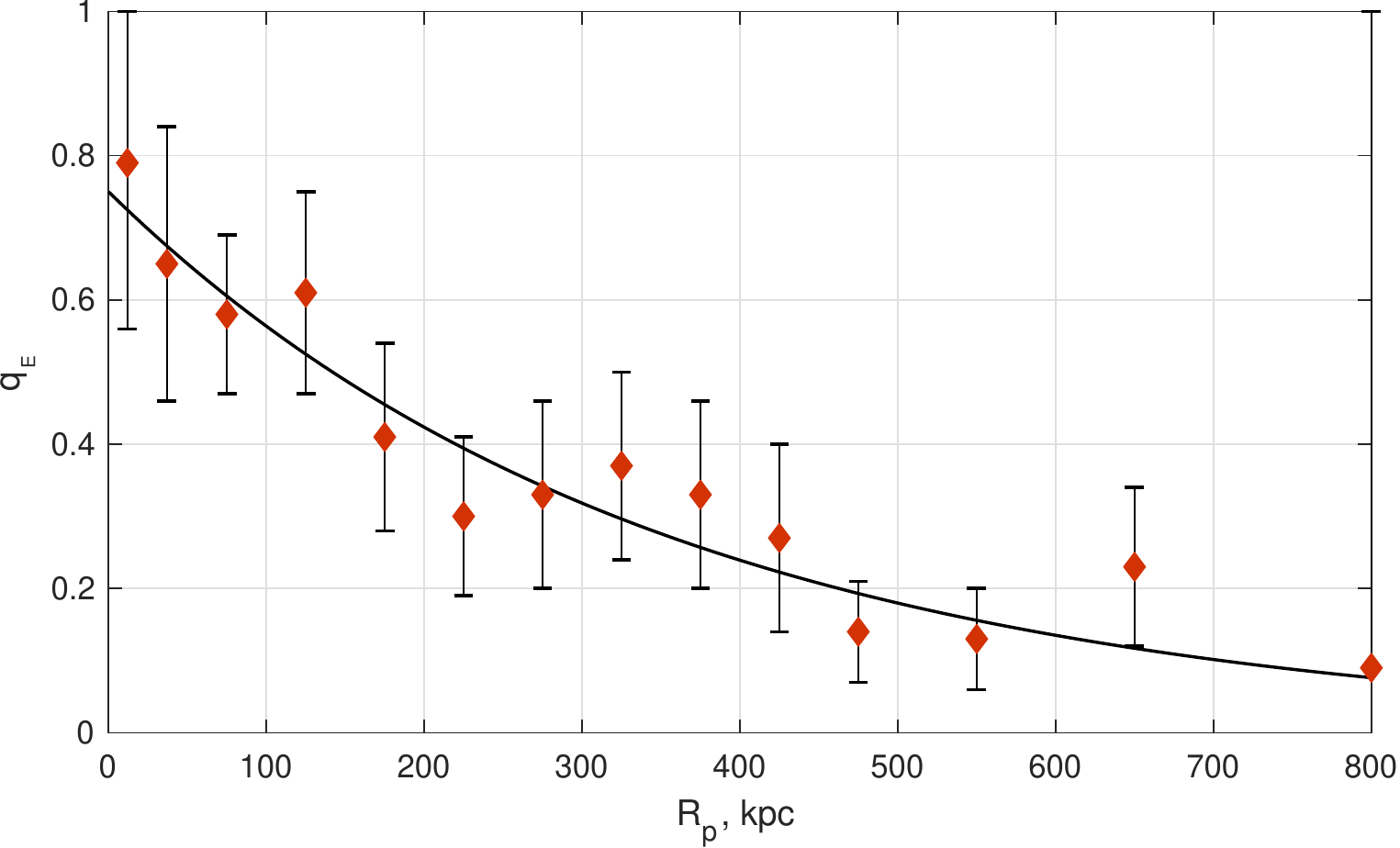}}
\caption{Relative number of dwarf satellites brighter than $L_K/L_{\odot} = 7.0$
dex around the massive Local Volume galaxies as a function of projected
separation.\label{fig7}}
\end{figure*}

The Local Volume galaxies with luminosities $L_K/L_{\odot}>10.5$ dex differ
significantly from each other by mean satellite system sizes $\langle R_p
\rangle$ and their radial velocities dispersions $\sigma_v$. The distribution of
these Milky Way-like galaxies by $\langle R_p\rangle$ and $\sigma_v$ is
presented in Figure~\ref{fig6}. The designations of early and late type galaxies
are the same as in previous figures. This diagram does not show any notable
correlation between the system size and velocity dispersion. The galaxies with
substantial bulges ($T<3$, red circles) have nearly the same mean separations of
their companions as the disk-dominated galaxies ($T>2$, blue circles). However,
the radial velocity dispersion of bulge-doninated galaxies is on the whole
higher than that of disc-dominated galaxies, according to their difference in
$M_T/L_K$ value. The evidence for that was provided also by Seo et al. (2020)
relying on richer data from  SDSS survey. Our Galaxy (plus sign) seems to be
untypical on this diagram. But, as we have already noted, $\sigma_v$ value for
the satellites of the Milky Way is slightly higher due to the inner position of
the observer and prevailing radial movements of the satellites. On the other
hand, the mean distance of the Milky Way satellites turnes out to be undervalued
because of the abundance of the ultra faint companions seen only at the closest
distances. So, the untypical behaviour of our Local group, reported by several
authors (Wang et al. 2021), is possibly caused by the selection effects.

According to Casertano \& van Gorkom (1991), Lucero et al. (2015) and Zobnina \&
Zasov (2020), the bright spirals in the Local Volume: NGC253, NGC2683, NGC2903,
NGC3521 and NGC5055 have declined (quasi-Keplerian) rotation curves at their
periphery. These five galaxies are marked in the Figure~\ref{fig6} with
strikethrough circles. Karachentsev et al. (2021) give for them mean values:
$\log(L_K/L_{\odot})=10.94, \sigma_v= 47$ km s$^{-1}, \langle R_p\rangle=248$
kpc, and $\log(M_T/M_{\odot})=11.66$. Their mean total mass-to-$K$ luminosity
ratio amounts $\langle M_T/L_K\rangle = 5.8\pm1.1$ in solar mass and luminosity
units. Such a low value of $M_T/L_K$ points out scanty dark halos around these
galaxies corresponding to the decrease of the rotation velocity at the periphery
of them. King \& Irwin (1997) noted  the declined rotation curve for one more
spiral galaxy, NGC3556. However, this galaxy has yet only one possible satellite
with coarsely measured radial velocity which makes its orbital mass estimate
quite indefinite. The existence of massive spiral galaxies with their
shallow	potential wells raises new unexpected questions with models for the
formation and evolution of stellar systems.

Using data on optical and HI rotation curves of spiral galaxies, Lapi et al.
(2018), Posti et al. (2019) and Sharma et al. (2021) noted that the
halo-mass-to-stellar mass ratio shows a tendency to decrease with increasing
stellar mass of the galaxy. This result, obtained on a scale of (1 -2) optical
radii of the galaxy, needs to be confirmed by the kinematics of satellites of
the host galaxies on the scale of the virial radius of their halos.

\section{Early and late type satellites around nearby massive galaxies.}\label{sec5}

While dwarf satellites move through the halo of massive galaxy they lose their
gas, which prevents the further star formation in them. This process transforms
gas-reach irregular (dIrr) and blue compact dwarfs (BCD) into spheroidal (dSph)
and elliptical dwarfs (dE). Obviously, the rate of this transformation increases
with closeness of a satellite to the main galaxy, with mass of the main galaxy
halo, and with shallowness of the potential well of the satellite itself. Dwarf
satellites are better represented in the Local Volume than in more distant
volumes, so we have used the existing data to test the expected patterns.

Dozens of ultra faint galaxies with luminosities $L_K\sim(10^2-10^5)L_{\odot}$
were detected within the Local Group (and cannot be observed in more distant
groups). To avoid selection effects we have limited ourselves by considering
dwarf galaxies with luminosities brighter than $1\times10^7L_{\odot}$. These
objects are thoroughly detectable at the far border of the Local Volume with
$D\simeq10$~Mpc. In the stacked group composed of 25 bright galaxies presented
in Table~\ref{tab1} there are 274 satellites more bright than the chosen limit;
118 from them, i.e. 43\%, are early type galaxies. The distribution of fraction
of the early type satellites along the stacked group radius is shown in
Figure~\ref{fig7} by the rhombs with error bars corresponding to the statistical
uncertainty. The behaviour of $q_E(R_p$) is well described by the exponential
law 
\begin{equation}
q_E(R_p)=0.75\times exp(-R_p/350),
\end{equation}
where $R_p$ is expressed in kpc. Thus, in the center of the stacked group the
relative number of satellites with suppressed star formation amounts to 75\%,
decreasing twice at the distance $(R_p)_{1/2}=242$~kpc. The value $(R_p)_{1/2}$
roughly corresponds to the halo virial radius of the massive galaxy with mean
luminosity of $L_K=8\times10^{10}L_{\odot}$. Such a matching is an independent
argument for a popular satellite quenching model in which quenched dwarfs are
formed while moving through massive neighbouring halo.

The data in Tables \ref{tab2}--\ref{tab5} show that the relative number of early
type satellites falls as the luminosity of the main galaxy decreases. Indeed,
the fraction of quenched satellites around the galaxies with luminosities
$\log(L_K/L_{\odot})= 9.5-10.5$ is 20\% (against 43\% for Milky Way-like
galaxies), while for LMC-like galaxies with $\log(L_K/L_{\odot})<9.5$ this
fraction amounts only to $q_E=10$\%. Hence, the mass of the main halo have a
significant impact on the gas abundance and stellar population age of the
satellites.

The expected trend of increasing the relative number of dSph and dE dwarfs among
less and less massive satellites can be barely tested by observations due to the
strong selection bias against faint dwarfs at larger distances. Generally, we
can note only that among 174 faint companions with luminosity $\log(L_K/
L_{\odot}) < 7.0$ around bright galaxies from Table~\ref{tab1} the relative
number of quenched dwarfs is $q_E=66$\%, i.e. 1.6 times higher than among
brighter companions. The more shallow potential wells of dwarfs facilitate
stripping.

In literature there is a large number of works discussing the abundance of
satellites around isolated galaxies of different mass and morphology based on
N-body simulations within the standard cosmological model (Brook et al. 2014,
Sales et al. 2013, Besla et al. 2018, Garrison-Kimmel et al. 2017, Wang et al.
2021, Font et al. 2021). To compare the simulation results and the observational
data, the common practice is to use the Local Group or to count the supposed
companions around SDSS galaxies. The first case raise the question as to how
typical is the Local Group itself. The second case is fraught with the problem
of separation of physical satellites from background and foreground galaxies.
The sample of 25 massive galaxies of the Local Volume, presented in
Table~\ref{tab1}, is indicative of diversity characterizing the neighboring
groups population.

\begin{figure}[t]
\centerline{\includegraphics[width=85mm]{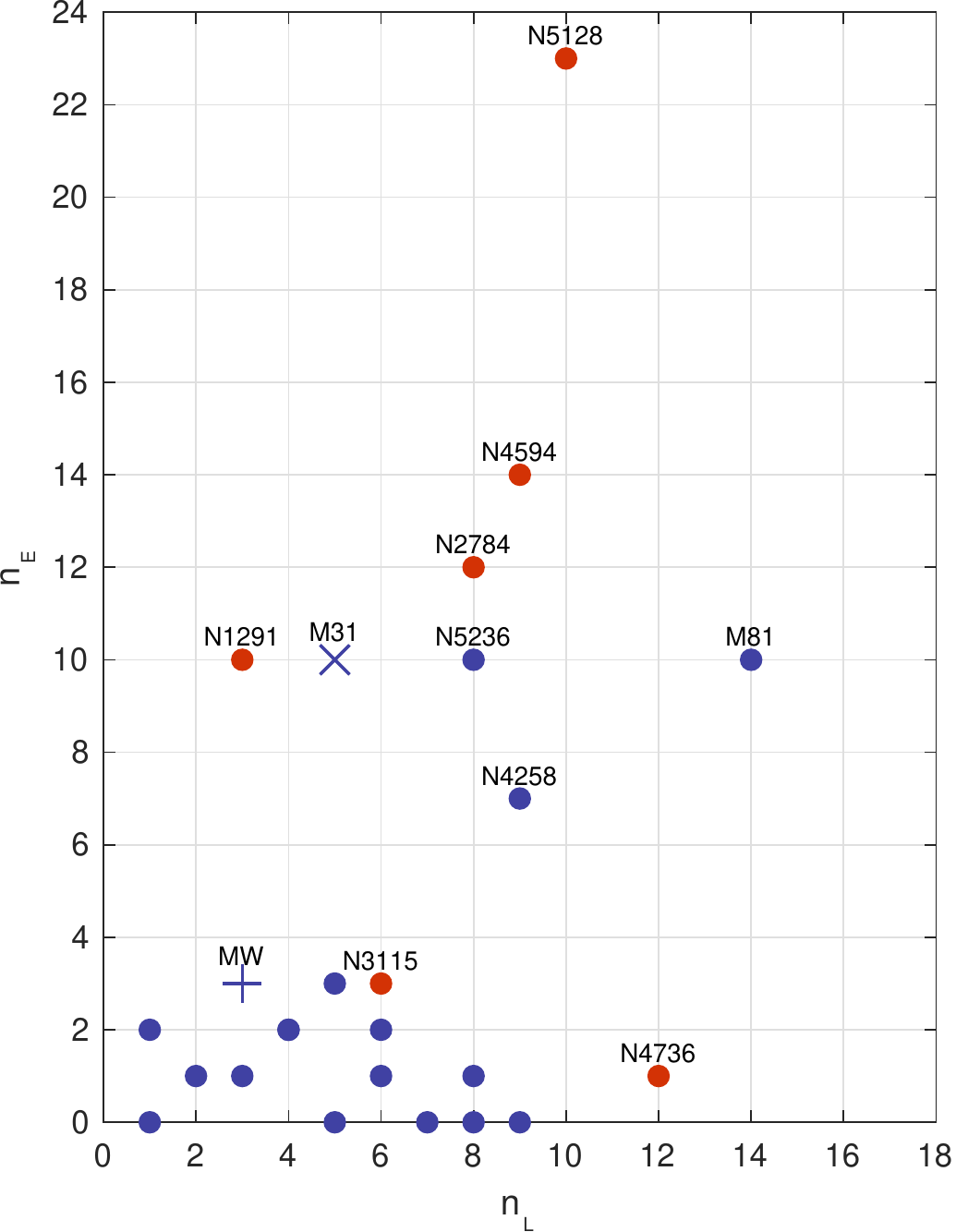}}
\caption{Distribution of bright Local Volume galaxies by the number of early and
late type satellites brighter than $L_K/L_{\odot}=7.0$ dex. The designations of
galaxies are the same as in Figure~\ref{fig4}.\label{fig8}}
\end{figure}

Figure~\ref{fig8} is a kind of passport for the local groups assembly. The
horizontal and the vertical axes correspond respectively to the number of late
$(n_L$) and early $(n_E$) type satellites around the main galaxies. To avoid the
selection of companions by their luminosity due to distance, we consider only
satellites brighter than $L_K/L_{\odot}=1\times 10^7$. Main galaxies of late and
early types are marked by blue and red circles. Milky Way and M31 are noted by
plus and cross signs.

As it can be seen from these data, there is a weak positive correlation between
numbers of late and early type satellites. The retinues of massive galaxies can
be divided conventionally into two classes: poor with $n_E\leq3$ and rich with
$n_E\geq7$. The first class comprises most retinues around main galaxies of late
type including the Milky Way retinue. Main galaxies with dominating bulges are
mostly presented in the second class, with M31 ranged among them. This trend is
consistent with Wang et al. (2021) stating that retinues of blue SDSS galaxies
more likely contain blue satellites because their dark halos are less massive
than the halos of red galaxies.

The data of Figure~\ref{fig8} can not imply that the satellite systems around
the Milky Way and M31 are atypical for the Local Volume galaxies.

It should be noted that the configuration of galaxies in the Figure~\ref{fig8}
may be revised when the outskirts of the neighboring massive galaxies are
studied consistently up to specified values of luminosity and surface brightness
of satellites, and within specified projected separation of about $\sim(1-2)$
virial radii of the main galaxy halo. The progress in measuring accurate
distances of satellites also can make some difference to the pattern of this
plot.

\section{Concluding remarks.}\label{sec6}

We used the data from the UNGC catalog (Karachentsev et al. 2013) to select
satellites around relatively isolated galaxies in the Local Volume. The sample
contains 476 supposed companions around 25 Milky Way-like galaxies with
luminosities of $L_K/L_{\odot}\geq10.5$ dex and 88 supposed companions around 47
main galaxies of smaller luminosities. About 64\% of these satellites have
measured radial velocities. Based on projected separations and radial velocities
of satellites, we determined the orbital masses (i.e. halo masses) of the
central galaxies. For the Milky Way-like galaxies the mean total mass-to-$K$
luminosity ratio is $\langle M_T/L_K\rangle=31.3\pm6.3$ in the solar mass and
luminosity units. The mean $M_T/L_K$ ratio shows the systematical grow towards
low luminosity galaxies. Early type galaxies (E, S0, Sa) have the total mass per
$K$ luminosity unit four times higher than late type spirals (Sb, Sc, Sd), being
respectively $\langle M_T/L_K\rangle=73\pm15(M_{\odot} / L_{\odot})$ and
$17.4\pm2.8(M_{\odot}/L_{\odot})$. The $M_T/L_K$ ratio increases as the mean
stellar density produced by the neighboring galaxies within a sphere of 1 Mpc
grows. Also, the low $M_T/L_K$ ratio is more characteristic of the galaxies
having luminosities one or two orders higher than luminosities of their
satellites. The minimal radial velocity dispersion values $\sigma_v <
55$~km~s$^{-1}$ are found for 5 spirals of the Local Volume: NGC253, NGC2683,
NGC2903, NGC3521 and NGC5055, all demonstrating declining rotation curves. Their
mean $M_T/L_K$ ratio is surprisingly small, $(5.8\pm1.1)M_{\odot}/L_{\odot}$,
and is comparable with the cosmic dark-to-baryonic matter ratio, $M_{\rm DM}/M_*
\simeq6$.

The fraction of early type satellites (dSph, dE) falls as the luminosity of the
central galaxy decreases and grows as the luminosities of the satellites
themselves decrease. For the stacked retinue of companions around the Milky
Way-like galaxies the relative number of quenched satellites amounts to
$\sim75$\% in the center, falling exponentially with the distance from the
parent galaxy with the decay constant of $\langle R_p\rangle=350$~kpc. These
characteristics of the quenched satellite subsystem are quite consistent with
the stripping model in which satellites lose their gas while moving through the
massive central halo.
 
Considering the satellites with luminosities $L_K/L_{\odot}>7.0$ dex around 25
bright isolated galaxies in the Local Volume with $L_K/L_{\odot}\geq10.5$ dex,
we conclude that our Galaxy and M31 galaxy seem to be quite typical in abundance
of satellites, in their morphology and kinematics.

Acknowledgements. This work is supported by the \fundingAgency{Russian
Scientific Foundation} grant No. \fundingNumber{19--02--00145}.

\nocite{*}
\bibliography{Karachentsev-Kashibadze}%

\begin{center}
\begin{table}[t]%
\centering
\caption{Sub-MW-like galaxies in the LV with $\log(L_K/L_{\odot}) = 10.0-10.5$ and their companions.\label{tab2}}%
\tabcolsep=0pt%
\begin{tabular*}{20pc}{@{\extracolsep\fill}lrrlrrrr@{\extracolsep\fill}}
\toprule
\textbf{Name} & \textbf{T} & \textbf{D} & \textbf{meth} & $\boldsymbol{V_{LG}}$ &
    $\boldsymbol{\log L_K}$ & $\boldsymbol{R_p}$ & $\boldsymbol{\Delta V}$ \\
\midrule
NGC\,925         &    7 & 9.55 & trgb & 738 & 10.14 &   0 &      0\\
d0226$+$3325     &   10 & 9.55 & mem  & 705 &  7.34 &  28 &  $-$33\\
AGC\,124640      &   10 & 9.55 & mem  & 834 &  7.14 &  63 &     96\\
DDO\,25          &    8 & 9.55 & mem  & 785 &  8.89 & 210 &     47\\
DDO\,19          &   10 & 9.55 & mem  & 771 &  8.03 & 422 &     33\\
UGCA\,127        &    6 & 8.50 & TF   & 562 & 10.00 &   0 &      0\\
UGCA\,127 sat    &   10 & 8.50 & mem  & 538 &  8.76 &  24 &  $-$24\\
HIPASSJ0622$-$07 &    8 & 8.40 & TF   & 591 &  9.45 & 108 &     29\\
NGC\,2787        &    1 & 7.48 & sbf  & 842 & 10.19 &   0 &      0\\
UGC\,4998        &    9 & 8.24 & trgb & 770 &  8.71 & 128 &  $-$72\\
NGC\,3344        &    4 & 9.82 & trgb & 500 & 10.33 &   0 &      0\\
NGC\,3344 dw1    &   10 & 9.82 & mem  & --- &  6.29 &  34 &    ---\\
NGC\,4490        &    7 & 8.91 & mem  & 623 & 10.28 &   0 &      0\\
NGC\,4485        &    8 & 8.91 & trgb & 517 &  8.99 &   9 & $-$106\\
MAPS1231$+$42    &   10 & 8.91 & mem  & 602 &  7.06 &  72 &  $-$21\\
$[$KK98$]$\,149  &   10 & 8.51 & trgb & 445 &  8.15 &  98 & $-$178\\
DDO\,129         &    8 & 8.90 & TF   & 582 &  8.81 & 197 &  $-$41\\
NGC\,4517        &    7 & 8.36 & trgb & 978 & 10.27 &   0 &      0\\
dw\,1232$+$0015  & $-$1 & 8.36 & mem  &1036 &  7.67 &  22 &     58\\
CGCG\,014$-$054  &    9 & 9.60 & TF   & 954 &  8.01 & 238 &  $-$24\\
NGC\,4559        &    6 & 8.91 & trgb & 787 & 10.20 &   0 &      0\\
AGC\,220847      &    9 & 9.18 & TF   & 777 &  7.72 & 264 &  $-$10\\
NGC\,4631        &    7 & 7.35 & trgb & 581 & 10.49 &   0 &      0\\
NGC\,4627        & $-$3 & 7.30 & mem  & 541 &  9.50 &   5 &  $-$40\\
NGC\,4631 dw2    &   10 & 7.30 & mem  & --- &  6.62 &  10 &    ---\\
NGC\,4631 dw3    & $-$2 & 7.30 & mem  & --- &  6.84 &  23 &    ---\\
BTS\,151         & $-$2 & 7.35 & sbf  & 734 &  7.75 &  36 &    153\\
PGC\,100707      &   10 & 7.30 & mem  & 693 &  7.45 &  41 &    112\\
NGC\,4631 dw1    &   10 & 7.30 & mem  & --- &  7.30 &  46 &    ---\\
HSC-9            & $-$1 & 7.05 & sbf  & --- &  7.07 &  47 &    ---\\
KDG\,178         &   10 & 7.30 & mem  & 763 &  7.16 &  55 &    162\\
HSC-10           & $-$1 & 7.05 & sbf  & --- &  6.83 &  74 &    ---\\
$[$KK98$]$\,141  &   10 & 7.11 & trgb & 568 &  6.98 & 542 &  $-$13\\
NGC\,4826        &    2 & 4.41 & trgb & 365 & 10.49 &   0 &      0\\
$[$KK98$]$\,177  & $-$2 & 4.83 & trgb & 228 &  7.42 & 132 & $-$137\\
UGC\,7929        &    9 & 4.40 & mem  & 334 &  7.01 & 205 &  $-$31\\
$[$KK98$]$\,180  & $-$2 & 4.40 & mem  & --- &  7.36 & 335 &    ---\\
DDO\,154         &   10 & 4.04 & trgb & 354 &  7.59 & 426 &  $-$11\\
IC\,5201         &    6 & 8.80 & TF   & 893 & 10.01 &   0 &      0\\
AM\,2220-460     &   10 & 8.80 & mem  & 822 &  7.62 &  78 &  $-$71\\

\bottomrule
\end{tabular*}
\end{table}
\end{center}

\begin{center}
\begin{table}[t]%
\centering
\caption{M33-like galaxies in the LV with $\log(L_K/L_{\odot}) = 9.5-10.0$ and their companions.\label{tab3}}%
\tabcolsep=0pt%
\begin{tabular*}{20pc}{@{\extracolsep\fill}lrrlrrrr@{\extracolsep\fill}}
\toprule
\textbf{Name} & \textbf{T} & \textbf{D} & \textbf{meth} & $\boldsymbol{V_{LG}}$ &
    $\boldsymbol{\log L_K}$ & $\boldsymbol{R_p}$ & $\boldsymbol{\Delta V}$ \\
\midrule
M\,33             &    6 &  0.93 & trgb &  34 &  9.62 &   0 &      0\\
And XXII          & $-$3 &  0.79 & trgb &  87 &  5.28 &  47 &     53\\
NGC\,1313         &    7 &  4.31 & trgb & 264 &  9.57 &   0 &      0\\
$[$KK98$]$\,27    &   10 &  4.23 & trgb & 327 &  7.04 &  25 &     63\\
NGC\,2283         &    6 & 10.0  & TF   & 622 &  9.84 &   0 &      0\\
KKSG\,9           &    9 & 10.0  & mem  & 474 &  8.72 &  63 & $-$148\\
IC\,2171          &    8 &  9.90 & TF   & 572 &  9.05 &  77 &  $-$50\\
NGC\,2403         &    6 &  3.19 & trgb & 262 &  9.86 &   0 &      0\\
MADCASHJ0742$+$65 & $-$2 &  3.39 & trgb & --- &  5.86 &  36 &    ---\\
DDO\,44           & $-$3 &  3.21 & trgb & 356 &  7.78 &  73 &     94\\
NGC\,2366         &    9 &  3.28 & trgb & 251 &  8.70 & 206 &  $-$11\\
ESO\,495$-$021    &    9 &  8.24 & trgb & 581 &  9.56 &   0 &      0\\
ESO\,496$-$010    &    9 &  8.59 & trgb & 524 &  8.43 & 415 &  $-$57\\ 
CGMW2$-$3473      &    9 &  7.40 & TF   & 549 &  8.31 & 431 &  $-$32\\
NGC\,3432         &    8 &  9.20 & TF   & 587 &  9.64 &   0 &      0\\
LV\,J1052$+$3628  &   10 &  9.20 & mem  & 451 &  7.04 &   8 & $-$136\\
UGC\,5983         &   10 &  9.20 & mem  & 741 &  7.80 &  10 &    154\\
LV\,J1052$+$3639  &   10 &  9.20 & mem  & 487 &  7.88 &  27 & $-$100\\
VV\,747           &   10 &  9.20 & mem  & 594 &  8.02 & 180 &      7\\
NGC\,4236         &    8 &  4.41 & trgb & 157 &  9.61 &   0 &      0\\
$[$KK98$]$\,125   &   10 &  4.40 & mem  & --- &  6.61 &  50 &    ---\\
DDO\,165          &    9 &  4.83 & trgb & 196 &  8.23 & 374 &     39\\
NGC\,4449         &    8 &  4.27 & trgb & 249 &  9.68 &   0 &      0\\
LV\,J1228$+$4358  & $-$1 &  4.07 & trgb & 273 &  8.53 &  11 &     24\\
NGC\,4656         &    8 &  7.98 & trgb & 635 &  9.93 &   0 &      0\\
NGC\,4656UV       &   10 &  7.98 & mem  & 568 &  8.88 &  40 &  $-$67\\
NGC\,5068         &    6 &  5.15 & trgb & 469 &  9.73 &   0 &      0\\
dw\,1318$-$21     & $-$1 &  4.90 & mem  & --- &  7.06 &  78 &    ---\\
NGC\,6503         &    6 &  6.25 & trgb & 309 & 10.00 &   0 &      0\\
$[$KK98$]$\,242   &   10 &  6.47 & trgb & --- &  6.44 &  31 &    ---\\
NGC\,7640         &    5 &  8.43 & trgb & 668 &  9.77 &   0 &      0\\
UGC\,12588        &    8 &  8.43 & mem  & 723 &  8.83 & 103 &     55\\
DDO\,217          &    8 &  8.55 & TF   & 720 &  9.37 & 219 &     52\\   
NGC\,7793         &    6 &  3.63 & trgb & 250 &  9.70 &   0 &      0\\
PGC\,704814       &   10 &  3.60 & mem  & 299 &  6.90 &  14 &     49\\

\bottomrule
\end{tabular*}
\end{table}
\end{center}

\begin{center}
\begin{table}[t]%
\centering
\caption{LMC-like galaxies in the LV with $\log(L_K/L_{\odot}) = 9.0-9.5$ and their companions.\label{tab4}}%
\tabcolsep=0pt%
\begin{tabular*}{20pc}{@{\extracolsep\fill}lrrlrrrr@{\extracolsep\fill}}
\toprule
\textbf{Name} & \textbf{T} & \textbf{D} & \textbf{meth} & $\boldsymbol{V_{LG}}$ &
    $\boldsymbol{\log L_K}$ & $\boldsymbol{R_p}$ & $\boldsymbol{\Delta V}$ \\
\midrule
NGC\,24            &    5 &  7.31 & trgb &   606 & 9.48 &   0 &      0\\
dw\,0009$-$25      & $-$1 &  7.31 & mem  &   --- & 6.00 &  14 &    ---\\
dw\,0010$-$25      & $-$1 &  7.31 & mem  &   --- & 7.39 &  52 &    ---\\
NGC\,404           & $-$1 &  2.98 & trgb &   193 & 9.26 &   0 &      0\\
Do-I               & $-$2 &  3.31 & trgb &   --- & 6.48 &  62 &    ---\\
NGC\,1156          &    8 &  6.98 & trgb &   507 & 9.21 &   0 &      0\\
NGC\,1156 dw1      &   10 &  6.98 & mem  &   --- & 6.55 &  17 &    ---\\
NGC\,1156 dw2      &   10 &  6.98 & mem  &   --- & 6.41 &  22 &    ---\\
LV\,J0300$+$2546   &   10 &  6.98 & mem  &   443 & 7.24 &  71 &  $-$64\\
dw\,0301$+$2446    &   10 &  6.98 & mem  &   --- & 6.57 &  74 &    ---\\
AGC\,124056        &   10 &  6.98 & mem  &   538 & 6.57 & 186 &     31\\
NGC\,1569          &    8 &  3.19 & trgb &   106 & 9.40 &   0 &      0\\
UGCA\,92           &   10 &  3.22 & trgb &    93 & 8.04 &  69 &  $-$13\\
NGC\,1744          &    7 & 10.00 & TF   &   574 & 9.42 &   0 &      0\\
ESO\,486$-$021     &    9 & 10.00 & mem  &   664 & 8.74 & 167 &     90\\
LMC                &    9 &  0.05 & cep  &    28 & 9.42 &   0 &      0\\
SMC                &    9 &  0.06 & cep  & $-$22 & 8.85 &  18 &  $-$50\\
NGC\,2337          &    9 & 11.86 & trgb &   476 & 9.34 &   0 &      0\\
UGC\,3698          &   10 & 11.27 & trgb &   464 & 8.36 &  37 &  $-$12\\
NGC\,2337 dw\,TBG1 &    9 & 11.9  & mem  &   --- & 7.14 &  72 &    ---\\
NGC\,4214          &    8 &  2.88 & trgb &   295 & 9.00 &   0 &      0\\
KDG\,90            & $-$2 &  2.98 & trgb &   --- & 7.59 &   9 &    ---\\
NGC\,4190          &    9 &  2.83 & trgb &   239 & 7.90 &  25 &  $-$56\\
NGC\,4163          &    9 &  2.99 & trgb &   163 & 7.92 &  36 & $-$132\\
MADCASH-2          &   10 &  3.00 & trgb &   --- & 5.80 &  70 &    ---\\
NGC\,5474          &    8 &  6.98 & trgb &   424 & 9.21 &   0 &      0\\
PGC\,2448110       &   10 &  6.98 & mem  &   392 & 7.02 &   4 &  $-$32\\

\bottomrule
\end{tabular*}
\end{table}
\end{center}

\begin{center}
\begin{table}[t]%
\centering
\caption{Dwarf galaxies with $\log(L_K/L_{\odot}) < 9.0$ in the LV and their companions.\label{tab5}}%
\tabcolsep=0pt%
\begin{tabular*}{20pc}{@{\extracolsep\fill}lrrlrrrr@{\extracolsep\fill}}
\toprule
\textbf{Name} & \textbf{T} & \textbf{D} & \textbf{meth} & $\boldsymbol{V_{LG}}$ &
    $\boldsymbol{\log L_K}$ & $\boldsymbol{R_p}$ & $\boldsymbol{\Delta V}$ \\
\midrule
NGC\,185             & $-$3 &  0.66 & trgb &  73 & 8.36 &   0 &     0\\
NGC\,147             & $-$3 &  0.76 & trgb &  85 & 8.21 &  11 &    12\\
UGC\,2716            &    8 &  6.66 & trgb & 467 & 8.36 &   0 &     0\\
UGC\,2684            &   10 &  7.08 & trgb & 438 & 7.57 & 112 & $-$29\\
ESO\,121$-$020       &   10 &  6.08 & trgb & 319 & 7.78 &   0 &     0\\
LV\,J0616$-$5745     &   10 &  6.1  & mem  & 286 & 7.07 &   6 & $-$33\\
DDO\,47              &    8 &  8.17 & trgb & 160 & 8.74 &   0 &     0\\
$[$KK98$]$\,65       &   10 &  7.98 & trgb & 170 & 8.11 &  40 &    10\\
CGCG\,262$-$028      &    9 &  9.80 & TF   & 497 & 8.18 &   0 &     0\\
KKH\,40              &   10 &  9.25 & trgb & 510 & 7.65 &  25 &    13\\
UGC\,5186            &   10 &  9.40 & TF   & 496 & 7.71 &   0 &     0\\
AGC\,198691          &   10 & 10.91 & NAM  & 464 & 6.42 &  37 & $-$32\\
DDO\,64              &   10 & 10.91 & TF   & 461 & 8.41 &   0 &     0\\
$[$KK98$]$\,78       &   10 & 10.9  & mem  & 479 & 7.29 &   6 &    18\\
NGC\,3109            &    8 &  1.34 & trgb & 110 & 8.58 &   0 &     0\\
Antlia               &   10 &  1.37 & trgb &  66 & 6.50 &  29 & $-$44\\
Antlia B             &   10 &  1.29 & trgb &  82 & 6.09 &  78 & $-$28\\
$[$KK98$]$\,94       &   10 & 10.19 & trgb & 684 & 7.32 &   0 &     0\\
LeG 21               &   10 & 10.4  & mem  & 696 & 6.90 &   7 &    12\\
LV\,J1157$+$5638     &   10 &  8.75 & trgb & 514 & 7.33 &   0 &     0\\
LV\,J1157$+$5638 sat &   10 &  8.75 & mem  & --- & 5.84 &   5 &   ---\\
UGC\,7584            &    9 &  6.14 & mem  & 545 & 7.40 &   0 &     0\\
KKH\,80              &   10 &  6.14 & trgb & 542 & 7.14 &  21 &  $-$3\\
dw\,1243$-$42        & $-$1 &  3.60 & mem  & --- & 6.68 &   0 &     0\\
dw\,1243$-$42b       &   10 &  3.60 & mem  & --- & 6.26 &   1 &   ---\\
NGC\,4688            &    6 &  6.70 & TF   & 855 & 8.50 &   0 &     0\\
$[$KK98$]$\,164      &    8 &  6.71 & TF   & 910 & 7.85 &  13 &    55\\
dw\,1252$-$40        & $-$1 &  3.60 & mem  & --- & 7.37 &   0 &     0\\
dw\,1251$-$40        & $-$1 &  3.60 & mem  & --- & 6.26 &   4 &   ---\\
DDO\,161             &    8 &  6.03 & trgb & 545 & 8.74 &   0 &     0\\
UGCA\,319            &   10 &  5.75 & trgb & 555 & 8.02 &  32 &    10\\
DDO\,168             &   10 &  4.25 & trgb & 270 & 8.13 &   0 &     0\\
DDO\,167             &   10 &  4.25 & trgb & 230 & 7.19 &  34 & $-$40\\
DDO\,169             &   10 &  4.41 & trgb & 348 & 7.73 &   0 &     0\\
DDO\,169 NW          &   10 &  4.33 & trgb & 328 & 6.34 &   4 & $-$20\\

\bottomrule
\end{tabular*}
\end{table}
\end{center}

\begin{center}
\begin{table*}[t]%
\caption{Mean parameters of the primary galaxies and their suites in the LV.\label{tab6}}
\centering
\begin{tabular*}{500pt}{@{\extracolsep\fill}lrcccccc@{\extracolsep\fill}}
\toprule
$\boldsymbol{L_K}$ \textbf{of primary} & $\boldsymbol{N_{\rm sat}}$ &
    $\boldsymbol{\langle \Delta V\rangle}$ & $\boldsymbol{\log\langle L_K\rangle}$ &
    $\boldsymbol{\langle R_p\rangle}$ & $\boldsymbol{\langle \sigma _V\rangle}$ &
    $\boldsymbol{\log\langle M_T\rangle}$ & $\boldsymbol{\log\langle M_T/L_K\rangle}$ \\ 
\midrule
(10.9--11.4) dex & 106 & $+$25\,$\pm$\,14 & 11.02 & 307 & 119 & 12.46 & 1.39$+$.10$-$.13\\
(10.5--10.9) dex & 192 &  $+$4\,$\pm$\,9  & 10.73 & 260 & 100 & 12.24 & 1.58$+$.11$-$.14\\
(10.0--10.5) dex &  23 &  $-$7\,$\pm$\,20 & 10.28 & 151 &  85 & 11.82 & 1.54$+$.10$-$.15\\
(9.5--10.0) dex  &  16 &  $-$7\,$\pm$\,19 &  9.70 & 125 &  81 & 11.62 & 1.92$+$.13$-$.16\\
(9.0--9.5) dex   &   9 & $-$26\,$\pm$\,23 &  9.30 &  68 &  64 & 11.53 & 2.23$+$.18$-$.31\\
(7.0--9.0) dex   &  15 &  $-$6\,$\pm$\,7  &  8.17 &  30 &  28 & 10.45 & 2.28$+$.12$-$.16\\

\bottomrule
\end{tabular*}
\end{table*}
\end{center}

\clearpage
\appendix
\section{Data\label{app1}}
\begin{center}
\setlength{\tabcolsep}{1pt}


\topcaption{The data on distances, radial velocities and luminosities of 380 supposed
companions around the 23 most bright galaxies of the Local Volume besides the
Local Group.\label{tabapp}}
\tablefirsthead{%
\toprule
\textbf{Name} & \textbf{T} & \textbf{D} & \textbf{meth} & $\boldsymbol{V_{LG}}$ &
  $\boldsymbol{\log L_K}$ & $\boldsymbol{R_p}$ & $\boldsymbol{\Delta V}$ \\
\midrule}
\tablehead{%
\multicolumn{8}{l}{{\bfseries TABLE~\ref{tabapp}} Continued}\\
\\[-0.5em]
\toprule
\textbf{Name} & \textbf{T} & \textbf{D} & \textbf{meth} & $\boldsymbol{V_{LG}}$ &
  $\boldsymbol{\log L_K}$ & $\boldsymbol{R_p}$ & $\boldsymbol{\Delta V}$ \\
\midrule}
\tabletail{\midrule}
\tablelasttail{\bottomrule}

\begin{xtabular*}{20pc}{@{\extracolsep\fill}lrrlrrrr@{\extracolsep\fill}}
\textbf{NGC\,253} & \textbf{5} & \textbf{3.70} & \textbf{trgb} & \textbf{276} & \textbf{10.98} & \textbf{0} & \textbf{0}\\
Scl-MM-Dw2         & $-$2 &  3.12 & trgb &   --- &  7.36 &   53 &    ---\\
Scl-MM-Dw1         & $-$2 &  3.94 & trgb &   --- &  6.77 &   72 &    ---\\
LV\,J0055$-$2310   &   10 &  3.62 & trgb &   288 &  6.17 &  176 &     12\\
Sculptor SR        &   10 &  3.70 & mem  &   --- &  6.36 &  258 &    ---\\
DDO\,6             &   10 &  3.43 & trgb &   347 &  7.08 &  278 &     71\\
NGC\,247           &    7 &  3.71 & trgb &   210 &  9.50 &  293 &  $-$66\\
Sc\,22             & $-$3 &  4.29 & trgb &   --- &  7.15 &  349 &    ---\\
ESO\,540$-$032     &   10 &  3.63 & trgb &   285 &  6.83 &  351 &      9\\
KDG\,2             &   10 &  3.56 & trgb &   290 &  6.85 &  468 &     14\\
ESO\,349-031       &   10 &  3.21 & trgb &   234 &  7.12 &  814 &  $-$42\\
NGC\,7793          &    6 &  3.71 & trgb &   250 &  9.70 &  846 &  $-$26\\
\\
\textbf{NGC\,628 = M74} & \textbf{5} & \textbf{10.19} & \textbf{trgb} & \textbf{827} & \textbf{10.60} & \textbf{0} & \textbf{0}\\
NGC\,628 dwB       &   10 & 10.2  & mem  &   --- &  5.30 &   34 &    ---\\
NGC\,628 dwA       &   10 & 10.2  & mem  &   --- &  7.03 &   37 &    ---\\
UGC\,1171          &   10 & 10.2  & mem  &   906 &  8.09 &  132 &     79\\
DDO\,13            &   10 &  9.04 & bs   &   798 &  8.53 &  149 &  $-$29\\
AGC\,112503        &    9 & 10.2  & mem  &   909 &  7.15 &  155 &     82\\
NGC\,628 dw\,TBG   &   10 & 10.2  & mem  &   --- &  6.78 &  174 &    ---\\
AGC\,114027        &   10 &  9.90 & bTF  &   898 &  6.81 &  220 &     71\\
AGC\,112454        &    9 & 10.2  & mem  &   837 &  7.35 &  297 &     10\\
KDG\,10            &   10 &  7.87 & bs   &   953 &  7.66 &  297 &    126\\
UGC\,1056          &    9 & 10.2  & mem  &   774 &  8.54 &  375 &  $-$53\\
JKB\,142           &   10 & 10.5  & bTF  &   887 &  7.01 &  403 &     60\\
UGC\,1104          &    9 &  7.55 & bs   &   869 &  8.35 &  481 &     42\\
\\
\textbf{NGC\,891} & \textbf{3} & \textbf{9.95} & \textbf{trgb} & \textbf{736} & \textbf{10.98} & \textbf{0} & \textbf{0}\\
$[$TT2009$]$\,25   &   10 & 10.28 & trgb &   903 &  7.23 &   43 &    167\\
$[$TT2009$]$\,30   &   10 &  9.95 & mem  &   --- &  6.79 &   62 &    ---\\
UGC\,1807          &    8 &  9.95 & mem  &   840 &  7.89 &   83 &    104\\
DDO\,24            &    8 &  9.95 & mem  &   780 &  8.97 &  128 &     44\\
UGC\,2172          &   10 &  9.30 & TF   &   765 &  8.32 &  649 &     29\\
DDO\,22            &   10 &  9.95 & mem  &   765 &  8.07 &  724 &     29\\
\\
\textbf{NGC\,1291} & \textbf{1} & \textbf{9.08} & t\textbf{rgb} & \textbf{702} & \textbf{10.97} & \textbf{0} & \textbf{0}\\
NGC\,1291 Dw3      & $-$1 &  9.08 & mem  &   --- &  7.84 &   24 &    ---\\
NGC\,1291 Dw6      & $-$1 &  9.08 & mem  &   --- &  7.46 &   49 &    ---\\
NGC\,1291 Dw2      & $-$1 &  9.08 & mem  &   --- &  7.60 &   98 &    ---\\
NGC\,1291 Dw8      & $-$1 &  9.08 & mem  &   --- &  8.10 &   98 &    ---\\
NGC\,1291 Dw5      & $-$1 &  9.08 & mem  &   --- &  7.72 &   98 &    ---\\
NGC\,1291 Dw4      & $-$1 &  9.08 & mem  &   --- &  7.81 &   98 &    ---\\
NGC\,1291 Dw13     & $-$1 &  9.08 & mem  &   --- &  7.28 &  114 &    ---\\
NGC\,1291 Dw15     &   10 &  9.08 & mem  &   --- &  6.30 &  133 &    ---\\
NGC\,1291 Dw12     & $-$1 &  9.08 & mem  &   --- &  6.52 &  184 &    ---\\
NGC\,1291 Dw1      & $-$1 &  9.08 & mem  &   --- &  7.38 &  185 &    ---\\
$[$KK2000$]$\,08   &    9 &  9.08 & mem  &   --- &  7.64 &  195 &    ---\\
ESO\,300$-$014     &    8 &  9.80 & TF   &   824 &  9.30 &  230 &    122\\
NGC\,1291 Dw9      & $-$1 &  9.08 & mem  &   --- &  7.04 &  241 &    ---\\
NGC\,1291 Dw10     & $-$1 &  9.08 & mem  &   --- &  7.58 &  260 &    ---\\
NGC\,1291 Dw14     &   10 &  9.08 & mem  &   --- &  6.60 &  274 &    ---\\
ESO\,300$-$4016    &   10 &  9.61 & trgb &   583 &  7.94 &  279 & $-$119\\
NGC\,1291 Dw11     &   10 &  9.08 & mem  &   --- &  6.86 &  312 &    ---\\
\\
\textbf{IC\,342} & \textbf{6} & \textbf{3.28} & \textbf{cep} & \textbf{244} & \textbf{10.60} & \textbf{0} & \textbf{0}\\
KK\,35             &   10 &  3.16 & trgb &   149 &  7.97 &   15 &  $-$95\\
UGCA\,86           &    8 &  2.98 & trgb &   280 &  9.13 &   89 &     36\\
KKH\,22            & $-$1 &  3.12 & trgb &   251 &  6.70 &  228 &      7\\
NGC\,1560          &    7 &  2.99 & trgb &   170 &  8.60 &  312 &  $-$74\\
NGC\,1569          &    8 &  3.19 & trgb &   106 &  9.40 &  312 & $-$138\\
Cam A              &   10 &  3.56 & trgb &   156 &  7.79 &  326 &  $-$78\\
Cam B              &   10 &  3.50 & trgb &   267 &  7.09 &  365 &     23\\
UGCA\,105          &    8 &  3.39 & trgb &   281 &  9.14 &  605 &     37\\
\\
\textbf{NGC\,2683} & \textbf{3} & \textbf{9.82} & \textbf{trgb} & \textbf{365} & \textbf{10.81} & \textbf{0} & \textbf{0}\\
NGC\,2683 dw1      &   10 &  9.82 & mem  &   334 &  6.49 &   34 &  $-$31\\
KK\,69             &   10 &  9.16 & trgb &   418 &  7.27 &   65 &     53\\
NGC\,2683 dw2      & $-$2 &  9.82 & mem  &   --- &  7.14 &   66 &    ---\\
KK\,70             & $-$3 &  9.29 & trgb &   --- &  7.86 &  102 &    ---\\
\\
\textbf{NGC\,2784} & \textbf{$-$2} & \textbf{9.82} & \textbf{sbf} & \textbf{398} & \textbf{10.80} & \textbf{0} & \textbf{0}\\
NGC\,2784 dw1      & $-$3 &  9.82 & mem  &   --- &  8.41 &    7 &    ---\\
KK\,73             & $-$1 &  9.82 & mem  &   --- &  9.08 &   16 &    ---\\
KSP-Dw18           &    9 &  9.82 & mem  &   --- &  6.58 &   24 &    ---\\
KK\,72             & $-$3 &  9.82 & mem  &   --- &  7.83 &   35 &    ---\\
KSP-Dw15           & $-$2 &  9.82 & mem  &   --- &  7.83 &   53 &    ---\\
KSP-Dw14           & $-$2 &  9.82 & mem  &   --- &  7.84 &   80 &    ---\\
KSP-Dw13           & $-$2 &  9.82 & mem  &   --- &  7.86 &   92 &    ---\\
KSP-Dw20           & $-$2 &  9.82 & mem  &   --- &  7.45 &   94 &    ---\\
KSP-Dw17           & $-$1 &  9.82 & mem  &   --- &  8.38 &  120 &    ---\\
KSP-Dw12           &   10 &  9.82 & mem  &   --- &  6.18 &  133 &    ---\\
KK\,74             & $-$1 &  9.82 & mem  &   --- &  8.52 &  137 &    ---\\
KSP-Dw24           &   10 &  9.82 & mem  &   --- &  6.58 &  171 &    ---\\
KSP-Dw16           &    9 &  9.82 & mem  &   --- &  6.66 &  176 &    ---\\
KSP-Dw23           & $-$2 &  9.82 & mem  &   --- &  6.32 &  178 &    ---\\
KSP-Dw21           & $-$2 &  9.82 & mem  &   --- &  6.45 &  180 &    ---\\
KK\,71             &   10 &  9.82 & mem  &   --- &  7.46 &  183 &    ---\\
DDO\,56            &   10 & 10.91 & TF   &   438 &  8.50 &  223 &     40\\
HIPASS J0916$-$23b &   10 &  9.82 & mem  &   550 &  8.47 &  223 &    152\\
KSP-Dw11           &    9 &  9.82 & mem  &   --- &  7.16 &  224 &    ---\\
KSP-Dw28           & $-$1 &  9.82 & mem  &   --- &  7.93 &  245 &    ---\\
KSP-Dw30           &   10 &  9.82 & mem  &   --- &  6.65 &  320 &    ---\\
ESO\,497$-$004     &    8 & 11.17 & TF   &   519 &  8.85 &  367 &    121\\
NGC\,2835          &    5 & 10.30 & TF   &   600 & 10.19 &  382 &    202\\
KSP-Dw6            & $-$2 &  9.82 & mem  &   --- &  6.99 &  390 &    ---\\
DDO\,62            &    8 & 10.86 & bTF  &   560 &  8.71 &  459 &    162\\
KSP-Dw3            & $-$2 &  9.82 & mem  &   --- &  8.22 &  560 &    ---\\
KSP-Dw4            &   10 &  9.82 & mem  &   --- &  7.72 &  566 &    ---\\
KSP-Dw1            & $-$1 &  9.82 & mem  &   --- &  7.80 &  626 &    ---\\
\\
\textbf{NGC\,2903} & \textbf{4} & \textbf{9.17} & \textbf{trgb} & \textbf{443} & \textbf{10.85} & \textbf{0} & \textbf{0}\\
UGC\,5086          & $-$3 &  8.70 & sbf  &   378 &  8.34 &   23 &  $-$65\\
NGC\,2903-HI       &   10 &  9.17 & mem  &   470 &  6.94 &   65 &     27\\
KDG\,56            &   10 &  9.17 & mem  &   441 &  7.44 &  245 &   $-$2\\
LSBC D565$-$06     &   10 &  9.29 & trgb &   386 &  7.34 &  460 &  $-$57\\
AGC\,198508        &   10 & 10.42 & trgb &   424 &  7.23 &  630 &  $-$19\\
\\
\textbf{M81 = NGC\,3031} & \textbf{3} & \textbf{3.70} & \textbf{trgb} & \textbf{104} & \textbf{10.95} & \textbf{0} & \textbf{0}\\
Holm IX            &   10 &  3.85 & trgb &   192 &  7.75 &   12 &     88\\
BK3N               &   10 &  4.17 & trgb &   101 &  6.12 &   12 &   $-$3\\
JKB3               &   10 &  3.70 & mem  &   198 &  5.63 &   17 &     94\\
A0952$+$69         &   10 &  3.93 & trgb &   242 &  6.87 &   18 &    138\\
Clump I            &   10 &  3.70 & mem  & $-$25 &  5.57 &   25 & $-$129\\
KDG\,61            & $-$1 &  3.66 & trgb &   360 &  8.11 &   32 &    256\\
KDG\,61em          &   10 &  3.70 & mem  &   255 &  5.98 &   32 &    151\\
D0959$+$68         &   10 &  4.27 & trgb & $-$46 &  6.44 &   35 & $-$150\\
Clump III          &   10 &  3.70 & mem  &    19 &  5.57 &   40 &  $-$85\\
M82                &    8 &  3.61 & trgb &   328 & 10.59 &   40 &    224\\
NGC\,3077          &    9 &  3.85 & trgb &   159 &  9.57 &   49 &     55\\
Garland            &   10 &  3.82 & trgb &   183 &  6.81 &   53 &     79\\
FM1                & $-$3 &  3.78 & trgb &   --- &  7.23 &   63 &    ---\\
d0944$+$69         & $-$3 &  3.84 & trgb &   --- &  5.88 &   64 &    ---\\
BK5N               & $-$3 &  3.70 & trgb &   --- &  7.18 &   73 &    ---\\
d1006$+$69         & $-$3 &  4.33 & trgb &   --- &  6.30 &   77 &    ---\\
IKN                & $-$3 &  3.75 & trgb &  $-$1 &  7.60 &   85 & $-$105\\
d1005$+$68         & $-$2 &  3.98 & trgb &   --- &  5.92 &   85 &    ---\\
d1009$+$68         & $-$3 &  3.73 & trgb &   --- &  6.25 &   85 &    ---\\
d0955$+$70         & $-$3 &  3.45 & trgb &   --- &  6.48 &   88 &    ---\\
NGC\,2976          &    7 &  3.66 & trgb &   142 &  9.44 &   89 &     38\\
KDG\,64            & $-$3 &  3.75 & trgb &   121 &  7.99 &  105 &     17\\
KK\,77             & $-$3 &  3.80 & trgb &   --- &  7.84 &  108 &    ---\\
d1015$+$69         & $-$1 &  4.07 & trgb &   --- &  6.06 &  114 &    ---\\
d1014$+$68         & $-$1 &  3.84 & trgb &   --- &  6.12 &  115 &    ---\\
F8D1               & $-$3 &  3.75 & trgb &     8 &  7.98 &  120 &  $-$96\\
d0934$+$70         & $-$3 &  3.02 & trgb &   --- &  6.64 &  137 &    ---\\
d1006$+$67         & $-$3 &  3.61 & trgb &   --- &  6.31 &  140 &    ---\\
d0958$+$66         &    9 &  3.82 & trgb &   221 &  7.12 &  145 &    117\\
HIJASS             &   11 &  3.70 & mem  &   187 &   --- &  150 &     83\\
Holm I             &   10 &  4.02 & trgb &   291 &  8.05 &  159 &    187\\
d0944$+$71         & $-$1 &  3.47 & trgb &   115 &  7.41 &  166 &     11\\
d0939$+$71         &   10 &  3.65 & trgb &   --- &  5.60 &  167 &    ---\\
KDG\,63            & $-$3 &  3.65 & trgb &     0 &  7.84 &  172 & $-$104\\
d0926$+$70         &   10 &  3.40 & trgb &   --- &  6.11 &  180 &    ---\\
HS\,117            &   10 &  3.96 & trgb &   116 &  6.72 &  195 &     12\\
IC\,2574           &    8 &  3.93 & trgb &   183 &  9.33 &  197 &     79\\
d1028$+$70         &   10 &  3.84 & trgb &    35 &  7.01 &  200 &  $-$69\\
DDO\,78            & $-$3 &  3.48 & trgb &   191 &  7.48 &  205 &     87\\
DDO\,82            &    9 &  3.93 & trgb &   207 &  8.39 &  219 &    103\\
d1041$+$70         & $-$2 &  3.70 & trgb &   --- &  6.36 &  265 &    ---\\
BK6N               & $-$3 &  3.31 & trgb &   --- &  7.25 &  316 &    ---\\
KDG\,73            &   10 &  3.91 & trgb &   263 &  6.60 &  330 &    159\\
UGC\,5497          &    9 &  3.73 & trgb &   267 &  7.19 &  340 &    163\\
KKH\,57            & $-$3 &  3.68 & trgb &   --- &  6.97 &  390 &    ---\\
UGC\,4483          &   10 &  3.58 & trgb &   304 &  7.09 &  520 &    167\\
DDO\,53            &   10 &  3.68 & trgb &   150 &  7.34 &  525 &     46\\
Holm II            &    9 &  3.47 & trgb &   311 &  9.20 &  545 &    207\\
\\
\textbf{NGC\,3115} & \textbf{$-$1} & \textbf{9.68} & \textbf{sbf} & \textbf{439} & \textbf{10.95} & \textbf{0} & \textbf{0}\\
KDG\,65            & $-$3 &  9.70 & mem  &   479 &  8.39 &   16 &     40\\
KKSG\,18           & $-$1 &  9.70 & mem  &   456 &  9.27 &   48 &     17\\
KKSG\,17           &   10 &  9.98 & trgb &   203 &  7.87 &  175 & $-$236\\
MCG-01-26-009      &   10 &  9.70 & mem  &   510 &  7.85 &  254 &     71\\
UGCA\,193          &    7 &  9.70 & mem  &   427 &  8.50 &  309 &  $-$12\\
KKSG\,16           & $-$3 &  9.70 & mem  &   --- &  7.85 &  360 &    ---\\
PGC\,154449        &    9 &  9.70 & mem  &   295 &  7.90 &  429 & $-$144\\
LV J0956$-$0929    &    9 &  9.38 & trgb &   378 &  7.95 &  471 &  $-$61\\
KKSG\,15           &   10 &  9.70 & mem  &   554 &  8.24 &  487 &    115\\
\\
\textbf{NGC\,3184} & \textbf{6} & \textbf{11.12} & \textbf{SN} & \textbf{588} & \textbf{10.52} & \textbf{0} & \textbf{0}\\
LV J1018$+$4109    & $-$1 & 11.12 & mem  &   --- &  7.71 &   50 &    ---\\
KUG1013$+$414      &    8 & 11.12 & mem  &   513 &  8.32 &   90 &  $-$75\\
SDSS1028$+$42      &   10 & 10.28 & NAM  &   551 &  7.33 &  445 &  $-$37\\
\\
\textbf{NGC\,3521} & \textbf{4} & \textbf{10.70} & \textbf{TF} & \textbf{598} & \textbf{11.09} & \textbf{0} & \textbf{0}\\
NGC\,3521sat       & $-$1 & 10.70 & mem  &   --- &  8.62 &   30 &    ---\\
KKSG\,20           &   10 & 10.70 & mem  &   636 &  7.37 &   56 &     38\\
NGC\,3521 dw\,TBG  & $-$2 & 10.70 & mem  &   --- &  7.28 &   72 &    ---\\
dw1110$+$0037      &   10 & 10.70 & mem  &   669 &  7.25 &  252 &     71\\
KKSG\,22           &   10 & 10.70 & mem  &   --- &  7.11 &  265 &    ---\\
UGC\,6145          &   10 & 10.70 & mem  &   546 &  7.83 &  341 &  $-$52\\
\\
\textbf{NGC\,3556 = M108} & \textbf{6} & \textbf{9.90} & \textbf{TF} & \textbf{777} & \textbf{10.52} & \textbf{0} & \textbf{0}\\
PGC\,034671        &    9 &  9.90 & mem  &   688 &  7.67 &  202 &  $-$89\\
\\
\textbf{NGC\,4258 = M106} & \textbf{4} & \textbf{7.66} & \textbf{trgb} & \textbf{506} & \textbf{10.92} & \textbf{0} & \textbf{0}\\
KDG\,101           & $-$1 &  7.28 & trgb &   330 &  8.41 &   29 & $-$176\\
$[$KKH2011$]$\,s11 & $-$3 &  7.66 & mem  &   --- &  6.55 &   43 &    ---\\
LV J1218$+$4655    &    8 &  8.28 & trgb &   477 &  7.49 &   53 &  $-$29\\
KK\,132            & $-$1 &  7.31 & trgb &   --- &  7.31 &   57 &    ---\\
M106 edgeN4217     & $-$2 &  7.70 & mem  &   --- &  7.39 &   67 &    ---\\
d1220$+$4649       & $-$3 &  7.90 & sbf  &   --- &  7.02 &   75 &    ---\\
BTS\,132           & $-$3 &  7.40 & sbf  &   --- &  7.36 &  119 &    ---\\
NGC\,4258 dwB      & $-$2 &  7.66 & mem  &   --- &  7.42 &  122 &    ---\\
NGC\,4288          &    7 &  8.24 & NAM  &   588 &  8.63 &  141 &     82\\
DDO\,120           &    9 &  7.28 & trgb &   516 &  8.73 &  206 &     10\\
NGC\,4258 dwC      & $-$2 &  7.66 & mem  &   --- &  7.15 &  208 &    ---\\
NGC\,4242          &    7 &  7.62 & trgb &   568 &  9.47 &  228 &     62\\
NGC\,4144          &    7 &  6.89 & trgb &   317 &  9.25 &  234 & $-$189\\
UGC\,7639          &    9 &  7.14 & sbf  &   450 &  8.33 &  249 &  $-$56\\
LV J1203$+$4739    &   10 &  7.94 & NAM  &   547 &  7.20 &  364 &     41\\
KK\,133            &   10 &  8.32 & NAM  &   601 &  7.04 &  525 &     95\\
KK\,151            &    9 &  8.20 & trgb &   479 &  7.79 &  651 &  $-$27\\
\\
\textbf{NGC\,4594 = M104} & \textbf{1} & \textbf{9.55} & \textbf{trgb} & \textbf{892} & \textbf{11.32} & \textbf{0} & \textbf{0}\\
Suc D1             & $-$1 &  9.55 & mem  &  1109 &  7.39 &    7 &    217\\
dw1240$-$1140      & $-$1 &  9.55 & mem  &  1097 &  7.18 &   15 &    205\\
KKSG\,32           & $-$1 &  9.00 & sbf  &   --- &  7.60 &   20 &    ---\\
dw1239$-$1143      & $-$1 &  7.90 & sbf  &  1171 &  8.27 &   37 &    279\\
Sombrero-DwA       & $-$1 &  9.70 & sbf  &   --- &  6.99 &   47 &    ---\\
Dw1241$-$1131      & $-$1 &  7.20 & sbf  &   --- &  6.98 &   47 &    ---\\
dw1239$-$1152      & $-$1 &  8.20 & sbf  &   --- &  6.16 &   53 &    ---\\
dw1240$-$1118      & $-$1 &  8.80 & sbf  &   --- &  8.53 &   53 &    ---\\
Sombrero-DwB       & $-$1 & 11.20 & sbf  &   --- &  7.20 &   67 &    ---\\
NGC\,4594-DGSAT-3  & $-$1 &  7.90 & sbf  &   --- &  7.39 &   68 &    ---\\
dw1239$-$1159      & $-$1 & 11.3  & sbf  &   --- &  7.32 &   70 &    ---\\
KKSG\,34           & $-$1 &  9.00 & sbf  &   --- &  7.80 &   73 &    ---\\
dw1237$-$1125      & $-$1 &  7.50 & sbf  &   --- &  7.57 &  118 &    ---\\
KKSG\,33           & $-$1 &  9.55 & mem  &   --- &  7.59 &  122 &    ---\\
PGC\,042730        & $-$1 &  9.55 & mem  &   829 &  8.72 &  171 &  $-$63\\
KKSG\,31           & $-$1 &  9.55 & mem  &   --- &  7.88 &  198 &    ---\\
PGC\,970397        &   10 & 10.0  & TF   &   928 &  8.04 &  201 &     36\\
KKSG\,29           &   10 &  9.82 & trgb &   562 &  7.67 &  220 & $-$330\\
PGC\,104868        &    9 & 11.17 & TF   &  1171 &  8.23 &  362 &    279\\
NGC\,4700          &    7 &  7.30 & TF   &  1219 &  8.86 &  372 &    327\\
KKSG\,30           &   10 &  9.72 & trgb &   918 &  7.71 &  471 &     26\\
UGCA\,307          &    9 &  9.03 & trgb &   731 &  8.31 &  573 & $-$161\\
NGC\,4802          &    0 & 11.5  & sbf  &   843 & 10.03 &  646 &  $-$49\\
PGC\,044460        &    8 &  8.70 & TF   &  1173 &  8.48 &  770 &    281\\
KKSG\,27           &    9 &  6.64 & bTF  &  1128 &  6.88 &  794 &    236\\
NGC\,4818          &    2 & 11.3  & TF   &   892 & 10.24 &  858 &      0\\
NGC\,4597          &    8 & 10.10 & TF   &   866 &  9.48 &  971 &  $-$26\\
\\
\textbf{NGC\,4736 = M94} & \textbf{2} & \textbf{4.41} & \textbf{trgb} & \textbf{352} & \textbf{10.56} & \textbf{0} & \textbf{0}\\
M94-Dw2            & $-$1 &  4.70 & trgb &   --- &  6.59 &   40 &    ---\\
M94-Dw1            & $-$1 &  4.10 & trgb &   --- &  6.75 &   73 &    ---\\
LV J1243$+$4127    &   10 &  4.81 & trgb &   444 &  6.77 &  105 &     92\\
KK\,160            &   10 &  4.33 & trgb &   346 &  6.60 &  218 &   $-$6\\
IC\,3687           &   10 &  4.57 & trgb &   377 &  8.19 &  238 &     25\\
IC\,4182           &    8 &  4.35 & trgb &   357 &  8.70 &  350 &      5\\
NGC\,4449          &    8 &  4.27 & trgb &   249 &  9.68 &  395 & $-$103\\
KK\,166            & $-$3 &  4.39 & trgb &   --- &  7.21 &  429 &    ---\\
CGCG\,189$-$050    &    9 &  3.93 & NAM  &   368 &  7.22 &  460 &     16\\
DDO\,126           &   10 &  4.97 & trgb &   230 &  8.09 &  470 & $-$122\\
DDO\,168           &   10 &  4.25 & trgb &   270 &  8.13 &  496 &  $-$82\\
UGC\,8215          &   10 &  4.57 & trgb &   304 &  7.18 &  501 &  $-$48\\
DDO\,167           &   10 &  4.25 & trgb &   230 &  7.19 &  510 & $-$122\\
NGC\,4244          &    6 &  4.31 & trgb &   259 &  9.52 &  560 &  $-$93\\
DDO\,169           &   10 &  4.41 & trgb &   348 &  7.73 &  600 &   $-$4\\
AGC\,239141        &   10 &  3.75 & NAM  &   359 &  6.01 &  602 &      7\\
NGC\,4395          &    8 &  4.76 & trgb &   308 &  9.47 &  704 &  $-$44\\
MCG+06-27-017      &    9 &  4.85 & trgb &   341 &  7.89 &  717 &  $-$11\\
\\
\textbf{NGC\,5055 = M63} & \textbf{4} & \textbf{9.04} & \textbf{trgb} & \textbf{570} & \textbf{11.00} & \textbf{0} & \textbf{0}\\
TBGdw2             &   10 &  9.04 & mem  &   --- &  6.51 &   57 &    ---\\
KK\,191            &   10 &  9.04 & mem  &   --- &  6.83 &   63 &    ---\\
UGC\,8313          &    8 &  9.04 & mem  &   658 &  8.46 &   65 &     88\\
KK\,193            &   10 &  9.04 & mem  &   --- &  6.71 &   85 &    ---\\
TBGdw5             & $-$1 &  9.04 & mem  &   --- &  6.54 &   87 &    ---\\
TBGdw3             & $-$1 &  9.04 & mem  &   --- &  6.26 &   88 &    ---\\
KKH\,82            &   10 &  7.48 & NAM  &   545 &  7.46 &   92 &  $-$25\\
TBGdw6             &   10 &  9.04 & mem  &   --- &  6.71 &   93 &    ---\\
TBGdw7             & $-$1 &  9.04 & mem  &   --- &  6.22 &   95 &    ---\\
TBGdw1             &   10 &  9.04 & mem  &   --- &  6.51 &  102 &    ---\\
TBGdw4             & $-$1 &  9.04 & mem  &   --- &  6.50 &  121 &    ---\\
CGCG\,217$-$018    &    9 &  9.04 & mem  &   608 &  8.19 &  256 &     38\\
dw1308$+$40        & $-$1 &  9.04 & mem  &   --- &  7.54 &  278 &    ---\\
dw1305$+$41        &   10 &  9.04 & mem  &   --- &  7.28 &  306 &    ---\\
dw1303$+$42        &   10 &  9.04 & mem  &   --- &  6.88 &  374 &    ---\\
KK\,194            &   10 &  9.04 & mem  &   --- &  7.15 &  387 &    ---\\
SDSS132753         &    9 &  6.58 & NAM  &   529 &  7.42 &  453 &  $-$41\\
\\
\textbf{NGC\,5128 = Cen-A} & \textbf{$-$2} & \textbf{3.68} & \textbf{trgb} & \textbf{310} & \textbf{10.89} &  \textbf{0} & \textbf{0}\\
KV19-329           & $-$1 &  3.68 & mem  &   382 &  7.03 &   25 &     72\\
KV19-271           & $-$1 &  3.68 & mem  &   302 &  7.12 &   31 &   $-$8\\
CenA-MM-Dw7        & $-$2 &  4.11 & trgb &   --- &  6.29 &   36 &    ---\\
$[$KK2000$]$\,55   & $-$3 &  3.85 & trgb &   282 &  7.81 &   43 &  $-$28\\
KK\,197            & $-$3 &  3.84 & trgb &   388 &  8.12 &   51 &     78\\
KV19-212           & $-$1 &  3.68 & mem  &   286 &  7.02 &   61 &  $-$24\\
KV19-442           & $-$1 &  3.68 & mem  &   242 &  7.11 &   64 &  $-$68\\
CenA-MM-Dw3        & $-$2 &  3.87 & trgb &   --- &  7.73 &   78 &    ---\\
CenA-MM-Dw4        & $-$2 &  4.09 & trgb &   --- &  6.95 &   84 &    ---\\
CenA-MM-Dw9        & $-$2 &  3.80 & trgb &   --- &  6.92 &   86 &    ---\\
CenA-MM-Dw2        & $-$2 &  4.15 & trgb &   --- &  6.17 &   91 &    ---\\
CenA-MM-Dw1        & $-$2 &  3.91 & trgb &   --- &  8.04 &   92 &    ---\\
CenA-MM-Dw5        & $-$2 &  3.61 & trgb &   --- &  5.59 &   93 &    ---\\
ESO\,324$-$024     &    8 &  3.78 & trgb &   272 &  8.35 &  102 &  $-$38\\
Fluffy             & $-$1 &  3.68 & mem  &   467 &  6.84 &  112 &    157\\
CenA-MM-Dw8        & $-$2 &  3.47 & trgb &   --- &  6.49 &  122 &    ---\\
dw1315-4309        & $-$2 &  3.70 & mem  &   --- &  5.74 &  127 &    ---\\
CenA-MM-Dw6        & $-$2 &  4.04 & trgb &   --- &  6.36 &  129 &    ---\\
dw1315-4232        & $-$2 &  3.70 & mem  &   --- &  6.32 &  131 &    ---\\
KK\,196            &   10 &  3.96 & trgb &   490 &  7.14 &  138 &    180\\
dw1323$-$40b       & $-$1 &  3.91 & trgb &   253 &  6.90 &  141 &    -57\\
dw1323$-$40        & $-$1 &  3.73 & trgb &   207 &  7.01 &  143 & $-$103\\
NGC\,5237          &    9 &  3.33 & trgb &   122 &  8.43 &  143 & $-$188\\
NGC\,5011C         & $-$1 &  3.73 & trgb &   394 &  7.98 &  145 &     84\\
dw1329$-$45        & $-$1 &  2.90 & trgb &   --- &  6.21 &  145 &    ---\\
dw1337$-$41        & $-$1 &  3.70 & mem  &   --- &  6.49 &  145 &    ---\\
dw1336$-$44        &   10 &  3.50 & trgb &   --- &  5.50 &  154 &    ---\\
KK\,189            & $-$3 &  4.23 & trgb &   501 &  7.48 &  169 &    191\\
dw1331$-$40        & $-$1 &  3.70 & mem  &   --- &  5.79 &  186 &    ---\\
ESO\,269$-$066     & $-$1 &  3.75 & trgb &   528 &  8.42 &  187 &    218\\
dw1341$-$43        & $-$1 &  3.53 & trgb &   398 &  6.86 &  191 &     88\\
$[$KK2000$]$\,57   & $-$3 &  3.84 & trgb &   275 &  7.28 &  193 &  $-$35\\
dw1342$-$43        & $-$1 &  2.90 & trgb &   273 &  6.61 &  204 &  $-$37\\
KK\,203            &   10 &  3.77 & trgb &    57 &  6.50 &  207 & $-$253\\
dw1322$-$39        &   10 &  2.95 & trgb &   413 &  6.12 &  207 &    103\\
KK\,213            & $-$3 &  3.77 & trgb &   --- &  7.52 &  217 &    ---\\
KK\,211            & $-$2 &  3.68 & trgb &   360 &  7.75 &  237 &     50\\
ESO\,325$-$011     &   10 &  3.40 & trgb &   311 &  7.86 &  244 &      1\\
KK\,217            & $-$3 &  3.50 & trgb &   --- &  7.22 &  295 &    ---\\
ESO\,269$-$058     &    9 &  3.75 & trgb &   140 &  8.86 &  307 & $-$170\\
dw1330$-$38        & $-$1 &  3.70 & mem  &   --- &  6.16 &  313 &    ---\\
$[$KK2000$]$\,53   & $-$3 &  2.92 & trgb &   --- &  7.46 &  317 &    ---\\
dw1302$-$40        & $-$1 &  3.70 & mem  &   --- &  6.59 &  318 &    ---\\
dw1257$-$41        & $-$1 &  3.70 & mem  &   --- &  7.13 &  330 &    ---\\
ESO\,269$-$037     &   10 &  3.15 & trgb &   481 &  6.96 &  333 &    171\\
NGC\,5206          & $-$3 &  3.21 & trgb &   334 &  8.95 &  345 &     24\\
dw1252$-$43        & $-$1 &  3.70 & mem  &   --- &  6.30 &  356 &    ---\\
KK\,221            &   10 &  3.82 & trgb &   268 &  6.74 &  368 &  $-$42\\
Cen-N              & $-$3 &  3.66 & trgb &   --- &  7.32 &  389 &    ---\\
NGC\,5102          &    1 &  3.66 & trgb &   227 &  9.70 &  415 &  $-$83\\
$[$KK2000$]$\,51   &   10 &  3.70 & mem  &   --- &  6.60 &  439 &    ---\\
dw1241$-$42        & $-$1 &  3.70 & mem  &   --- &  6.41 &  472 &    ---\\
NGC\,4945          &    6 &  3.47 & trgb &   299 & 10.66 &  474 &  $-$11\\
$[$KK2000$]$\,58   & $-$3 &  3.36 & trgb &   254 &  7.15 &  499 &  $-$56\\
dw1258$-$37        & $-$1 &  3.70 & mem  &   --- &  6.52 &  500 &    ---\\
dw1240$-$42        & $-$1 &  3.70 & mem  &   --- &  6.86 &  500 &    ---\\
ESO\,383$-$087     &    8 &  3.19 & trgb &   108 &  9.05 &  539 & $-$202\\
ESO\,219$-$010     & $-$3 &  4.29 & sbf  &   --- &  8.03 &  562 &    ---\\
ESO\,174$-$001     &   10 &  3.60 & TF   &   438 &  7.90 &  701 &    128\\
$[$KK2000$]$\,54   &   10 &  3.75 & trgb &   394 &  6.26 &  723 &     84\\
PGC\,051659        &   10 &  3.61 & trgb &   177 &  7.51 &  753 & $-$133\\
NGC\,5253          &    8 &  3.44 & trgb &   193 &  9.08 &  761 & $-$117\\
\\
\textbf{NGC\,5194 = M51} & \textbf{5} & \textbf{8.40} & \textbf{SN} & \textbf{538} & \textbf{10.97} & \textbf{0} & \textbf{0}\\
NGC\,5195          &    0 &  7.66 & sbf  &   548 & 10.59 &   12 &     10\\
NGC\,5229          &    7 &  8.49 & trgb &   359 &  8.51 &  151 & $-$179\\
dw1340$+$45        &   10 &  8.40 & mem  &   --- &  6.74 &  342 &    ---\\
LV J1328$+$4937    &   10 &  8.40 & mem  &   497 &  7.20 &  349 &  $-$41\\
LV J1342$+$4840    &    9 &  8.40 & mem  &   543 &  7.60 &  394 &      5\\
UGCA\,361          & $-$1 &  8.40 & mem  &   --- &  8.13 &  409 &    ---\\
MCG+08-25-028      &   10 &  8.40 & mem  &   565 &  7.77 &  438 &     32\\
dw1313$+$46        &   10 &  8.40 & mem  &   --- &  6.98 &  443 &    ---\\
dw1338$+$50        & $-$1 &  8.40 & mem  &   --- &  7.08 &  485 &    ---\\
dw1327$+$51        & $-$1 &  8.40 & mem  &   --- &  7.00 &  645 &    ---\\
\\
\textbf{NGC\,5236 = M83} & \textbf{5} & \textbf{4.90} & \textbf{trgb} & \textbf{307} & \textbf{10.86} & \textbf{0} & \textbf{0}\\
KK\,208            & $-$3 &  5.01 & trgb &   --- &  8.71 &   27 &    ---\\
UGCA\,365          &   10 &  5.42 & trgb &   367 &  7.73 &   55 &     60\\
dw1340$-$30        & $-$2 &  5.08 & trgb &   --- &  6.90 &   76 &    ---\\
NGC\,5264          &    8 &  4.79 & trgb &   269 &  8.88 &   86 &  $-$38\\
IC\,4316           &   10 &  4.35 & trgb &   369 &  8.21 &  104 &     62\\
ESO\,444$-$084     &   10 &  4.61 & trgb &   380 &  7.61 &  157 &     73\\
KK\,218            & $-$3 &  4.94 & trgb &   --- &  7.40 &  158 &    ---\\
dw1328$-$29        & $-$1 &  4.90 & mem  &   --- &  6.75 &  159 &    ---\\
dw1336$-$32        & $-$2 &  4.90 & mem  &   --- &  7.19 &  189 &    ---\\
IC\,4247           &   10 &  5.18 & trgb &   200 &  8.25 &  195 & $-$107\\
dw1326$-$29        & $-$2 &  4.90 & mem  &   --- &  6.88 &  201 &    ---\\
dw1330$-$32        &   10 &  4.90 & mem  &   --- &  6.04 &  222 &    ---\\
dw1329$-$32        &    9 &  4.90 & mem  &   --- &  6.73 &  244 &    ---\\
KK\,200            &   10 &  4.76 & trgb &   271 &  7.31 &  249 &  $-$36\\
Dw1337$-$26        & $-$2 &  4.90 & mem  &   --- &  7.08 &  268 &    ---\\
HIDEEP J1337$-$33  &   10 &  4.55 & trgb &   371 &  6.73 &  301 &     64\\
dw1337$-$33        &   10 &  4.90 & mem  &   --- &  6.52 &  302 &    ---\\
dw1325$-$33        & $-$1 &  4.90 & mem  &   --- &  6.78 &  305 &    ---\\
KK\,195            &   10 &  5.22 & trgb &   345 &  6.96 &  326 &     38\\
dw1321$-$27        & $-$1 &  4.90 & mem  &   --- &  6.79 &  335 &    ---\\
dw1322$-$27        & $-$1 &  4.90 & mem  &   --- &  7.17 &  335 &    ---\\
dw1341$-$33        & $-$2 &  4.90 & mem  &   --- &  7.06 &  342 &    ---\\
dw1357$-$28        & $-$1 &  4.90 & mem  &   --- &  6.50 &  357 &    ---\\
KK\,198            & $-$3 &  4.90 & mem  &   --- &  7.39 &  393 &    ---\\
dw1314$-$28        & $-$1 &  4.90 & mem  &   --- &  7.25 &  439 &    ---\\
dw1343$-$34        & $-$1 &  4.90 & mem  &   --- &  6.31 &  455 &    ---\\
dw1401$-$32        & $-$1 &  4.90 & mem  &   --- &  6.87 &  458 &    ---\\
dw1326$-$35        &   10 &  4.90 & mem  &   --- &  6.14 &  479 &    ---\\
dw1406$-$29        & $-$1 &  4.90 & mem  &   --- &  6.87 &  512 &    ---\\
dw1403$-$33        & $-$1 &  4.90 & mem  &   --- &  6.75 &  525 &    ---\\
dw1306$-$29        & $-$1 &  4.90 & mem  &   --- &  6.87 &  528 &    ---\\
ESO\,384$-$016     &    9 &  4.49 & trgb &   350 &  7.83 &  596 &     43\\
dw1301$-$30        & $-$1 &  4.90 & mem  &   --- &  6.72 &  601 &    ---\\
dw1409$-$33        & $-$1 &  4.90 & mem  &   --- &  6.88 &  613 &    ---\\
AM1320$-$230       & $-$3 &  4.90 & mem  &   --- &  7.46 &  619 &    ---\\
dw1415$-$32        &   10 &  4.90 & mem  &   --- &  6.08 &  680 &    ---\\
dw1413$-$34        & $-$1 &  4.90 & mem  &   --- &  6.26 &  695 &    ---\\
dw1410$-$34        &   10 &  4.90 & mem  &   --- &  6.54 &  701 &    ---\\
dw1318$-$21        & $-$1 &  4.90 & mem  &   --- &  7.06 &  781 &    ---\\
HIPASS J1337$-$39  &   10 &  5.08 & trgb &   258 &  7.21 &  870 &  $-$49\\
\\
\textbf{M101 = NGC\,5457} & \textbf{6} & \textbf{6.95} & \textbf{trgb} & \textbf{378} & \textbf{10.79} & \textbf{0} & \textbf{0}\\
NGC\,5477          &    9 &  6.77 & trgb &   451 &  8.26 &   44 &     73\\
M101-df1           &   10 &  6.37 & trgb &   --- &  6.17 &   53 &    ---\\
NGC\,5474          &    8 &  6.82 & trgb &   424 &  9.20 &   96 &     46\\
M101-df2           & $-$1 &  6.87 & trgb &   --- &  6.74 &   96 &    ---\\
M101-df3           & $-$1 &  6.52 & trgb &   --- &  7.27 &   97 &    ---\\
M101-DwA           & $-$1 &  6.64 & trgb &   --- &  6.97 &  102 &    ---\\
UGC\,8882          & $-$1 &  7.80 & sbf  &   482 &  7.64 &  110 &    104\\
GBT1355$+$54       &   11 &  6.95 & mem  &   345 &   --- &  152 &  $-$33\\
Holm IV            &    8 &  6.93 & trgb &   272 &  8.60 &  163 & $-$106\\
M101-Dw9           & $-$1 &  7.71 & trgb &   --- &  6.17 &  166 &    ---\\
NGC\,5585          &    7 &  7.00 & trgb &   457 &  9.07 &  429 &     79\\
dw1343$+$58        &    9 &  6.30 & TF   &   338 &  7.65 &  605 &  $-$40\\
DDO\,194           &    8 &  6.30 & trgb &   381 &  8.13 &  647 &      3\\
\\
\textbf{NGC\,6744} & \textbf{4} & \textbf{9.51} & \textbf{trgb} & \textbf{706} & \textbf{10.91} & \textbf{0} & \textbf{0}\\
$[$KK2000$]$\,71   &   10 &  9.50 & mem  &   --- &  8.24 &   30 &    ---\\
$[$KK2000$]$\,72   &   10 &  9.50 & mem  &   --- &  7.12 &   50 &    ---\\
$[$KK2000$]$\,70   &   10 &  9.50 & mem  &   --- &  7.21 &   55 &    ---\\
NGC\,6744 dw\,TBGb &   10 &  9.50 & mem  &   --- &  5.88 &   64 &    ---\\
ESO\,104$-$044     &   10 &  9.73 & trgb &   614 &  8.31 &   66 &  $-$92\\
NGC\,6744 dw\,TBGa & $-$1 &  9.50 & mem  &   --- &  7.38 &  120 &    ---\\
ESO\,104$-$022     &   10 &  8.95 & trgb &   654 &  8.05 &  302 &  $-$52\\
AM1909$-$615       &    8 &  9.50 & mem  &   821 &  7.92 &  333 &    115\\
NGC\,6684          &    0 &  8.70 & TF   &   720 & 10.39 &  435 &     14\\
IC\,4870           &    9 &  8.51 & trgb &   740 &  8.34 &  592 &     34\\
\\
\textbf{NGC\,6946} & \textbf{6} & \textbf{7.73} & \textbf{trgb} & \textbf{355} & \textbf{10.99} & \textbf{0} & \textbf{0}\\
KK\,251            &   10 &  7.73 & mem  &   433 &  7.94 &   77 &     78\\
UGC\,11583         &   10 &  7.30 & trgb &   429 &  8.16 &   86 &     74\\
KK\,252            &   10 &  7.73 & mem  &   441 &  8.13 &  105 &     86\\
KKR\,55            &   10 &  7.73 & mem  &   337 &  8.40 &  178 &  $-$18\\
KKR\,56            &   10 &  7.73 & mem  &   264 &  8.24 &  311 &  $-$91\\
Cepheus 1          &    8 &  7.73 & mem  &   342 &  9.61 &  525 &  $-$13\\
KKR\,59            &    8 &  7.73 & mem  &   307 &  9.39 &  631 &  $-$48\\
KKR\,60            &   10 &  7.73 & mem  &   296 &  8.66 &  672 &  $-$59\\

\end{xtabular*}
\begin{tablenotes}
\item  These data are also available online at the CDS via
   \url{http://cdsarc.u-strasbg.fr/viz-bin/cat/J/AN/...} 
\end{tablenotes}

\end{center}

\end{document}